\documentclass[12pt,preprint]{aastex}
\usepackage{graphics}

\shorttitle{Analytic Line Fitting}
\shortauthors{De~Vries \& Myers}

\renewcommand{\sec}{\mbox{$^{\prime\prime}$}}
\newcommand{\jequals}[2]{\mbox{$J = {#1}\rightarrow{#2}$}}
\newcommand{\nthp}{N$_{2}$H$^{+}$}
\newcommand{\hcop}{HCO$^{+}$}
\newcommand{\hthcop}{H$^{13}$CO$^{+}$}

\begin{document}
\sloppy
\title{Molecular Line Profile Fitting with Analytic Radiative Transfer Models}
\author{Christopher H. De Vries and Philip C. Myers}
\affil{Harvard-Smithsonian Center for Astrophysics}

\begin{abstract}
We present a study of analytic models of starless cores whose line profiles
have ``infall asymmetry,'' or blue-skewed shapes indicative of contracting
motions. We compare the ability of two types of analytical radiative transfer
models to reproduce the line profiles and infall speeds of centrally condensed
starless cores whose infall speeds are spatially constant and range between 0
and 0.2~km~s$^{-1}$. The model line
profiles of \hcop~(\jequals{1}{0}) and \hcop~(\jequals{3}{2}) are produced by a
self-consistent Monte Carlo radiative transfer code. The analytic models assume
that the excitation temperature in the front of the cloud is either constant
(``two-layer'' model) or increases inward as a linear function of optical depth
(``hill'' model). Each analytic model is matched to the line profile by rapid
least-squares fitting.

The blue-asymmetric line profiles with two peaks, or with a
blue shifted peak and a red shifted shoulder, can be well fit by one or both of
the analytic models. 
For two-peak profiles is
best matched by the ``{\sc hill5}'' model (a five parameter version of the hill
model), with an RMS error of
0.01~km~s$^{-1}$, while the ``{\sc twolayer6}'' model underestimates the infall
speed by a factor of $\sim 2$. For red-shoulder profiles, the {\sc hill5} and
{\sc twolayer6} fits reproduce infall speeds equally well, with an RMS error of
0.04~km~s$^{-1}$. The fits are most accurate when the line has a brightness
temperature greater than 3~K. Our most accurate models tend to
reproduce not only the line
profile shape, but also match the excitation conditions along the line of
sight. A better match to the line profile shape does not
necessarily imply a better match to the infall speed and provide guidance on
how to minimize the risk of obtaining a poor infall speed fit. 

A peak signal to noise ratio of at least 30 in the
molecular line observations is required for performing these analytic radiative
transfer fits to the line profiles. Moderate amounts of
depletion and beam smoothing do not adversely affect the accuracy of the infall
speeds obtained from these models.
\end{abstract}

\keywords{stars: formation --- radio lines: ISM --- radiative transfer}

\section{Introduction}
There are four distinct stages in isolated star
formation: (1) formation of a gravitationally bound core withing a molecular
cloud, (2) the gravitationally driven collapse of that core, (3) the formation
of a central star-like object, (4) dispersal of the remaining cloud material
\citep*{sal}. The initial two stages are
predominantly studied through millimeter- and submillimeter-wave-length
observations. The molecular line rotational transition emission visible in
these wavelength regimes allow one to probe both the physical and chemical
composition of the molecular cloud cores out of which stars form. Spectral
analysis of the molecular line emission also allows one to investigate the
kinematic motions within the molecular cloud, particularly the phase of
gravitational collapse \citep{meo} as well as the associated bipolar outflows
which are ubiquitous around young stars \citep{bac}.

In a dense cloud core, commonly used molecular line tracers of moderate optical
depth, such as CS~(\jequals{2}{1}), or \hcop~(\jequals{1}{0}) tend to become
self-absorbed. As the cloud contracts under the
influence of gravity, the central regions tend to become more dense than the
outer regions. This sets up a gradient in the excitation temperature of the
molecular gas within the cloud such that the excitation temperatures in the
central regions of the cloud are greater than those in the outer regions, even
if the kinetic temperature is constant within the molecular cloud. If the cloud
is spherical and static the spectral line emission from this cloud in these
tracers will be symmetric about the line of sight velocity and
self-absorbed. However, if there is radial motion around the molecular cloud
core, a velocity gradient develops around the center of the core, precisely
where the excitation temperature is the greatest. Much of the emission from the
rear 
of the cloud is not absorbed as it travels through the core, due to the high
excitation temperature within the core, and continues unabsorbed through the
front section of the cloud because the velocity gradient between the
rear and front sections of the cloud results in a Doppler shift which is
significant compared to the broadening by thermal motions. Emission from the
front of the cloud however is absorbed by
nearby molecular gas with lower excitation temperature and only a small
Doppler shift. In the case of inward radial motion the rear emission is
blue-shifted and the front emission is red-shifted, yielding the ``classic''
blue-asymmetric infall line profile. This line profile has become the
predominant tool for investigating infall motions within star-forming molecular
cloud cores \citep{sl,zhou2,zhou,mmtwbg,gemm,lmt2}. 

Although this
blue-asymmetry results from infall, it is not unique to that phenomenon. The
superposition of multiple clouds along the line of sight can also cause an
asymmetric line profile. By observing an optically thin molecular line, which
should remain symmetric around the line of sight velocity of the cloud core, it
is possible to
reduce the likelihood of mistaking separate clouds for an infalling
cloud. Asymmetric line profiles can also arise in kinematically complex
regions. Studies have shown that rotation and outflow can also produce
asymmetric line profiles \citep{slp,al,nmwb}. Even the asymmetric line emission
of B335, once
thought to be the prototypical 
example of spectrally detected inside-out collapse \citep{zekw,zekw2,zhou}, is
now also thought to also contain an outflow which gives rise to the line wings
previously modeled as part of the collapse
\citep{wmmt}. In spite of the multiple molecular cloud configurations that can
result in blue-asymmetric line profiles, they are still the best way to study
infall motions available, provided we understand the context in which they
occur and make clear the necessary caveats. 

Understanding how blue-asymmetric line profiles arise from infall motion is
easy, but creating a complete model matching the molecular emission from one or
more tracers is a difficult and time-consuming task. Typically one constructs
an infalling physical model and uses LVG, Monte Carlo \citep{cegw} or
microturbulent \citep{zhou} radiative transfer to model the emission. By
iteratively changing the model parameters one is able to match some or all of
the observations reasonably well. This process can take several minutes per
iteration, and becomes prohibitively slow as the number of physical parameters
becomes large and as the number of molecular species increases. The derived
cloud parameters are also highly dependent on
the input
physical model. Changing the physical model can lead to an entirely different
set of solutions.

``Starless cores'' are useful regions for observations of inward motion because
they are numerous, nearby, relatively simple in their density structure, and
their motions are not confused by outflows. They have been the subject of
several studies \citep{lm,lmt1,lmt2,bapabw,mm,tmcwc,tmcw,ts,plm}, some of which
have used radiative transfer models to fit 
their line profiles.
In this paper we present simple analytic radiative transfer
models, based on the simplest available assumptions of possible excitation
temperature trends within starless molecular cloud cores. These models can
easily be fit
to blue-asymmetric self-absorbed molecular line profiles, and produce an
estimate of the infall velocity suggested by those line profiles. We also test
the efficacy of the analytic radiative transfer models by comparing the infall
velocities derived using those models with actual infall velocities in
rigorously modeled condensing molecular clouds. Our main finding is that our
analytic radiative transfer models can yield good estimates for the physical
conditions of the starless cores with few assumptions and little computational investment.

\section{Analytic Models}
We compare two analytic models in this paper. Although it is possible to
construct more analytic models using the same techniques in this paper we feel
that these two models are the most appropriate for modeling infall. It is
possible to analytically integrate the equation of transfer
\begin{equation}
\frac{dT_{\rm B}}{d\tau_{\nu}} = -T_{\rm B} + J(T),
\end{equation}
where $T_{\rm B}$ is the brightness temperature, $T$ is the excitation
temperature and $J(T) =
\left(h\nu/k\right)\left[\exp\left(h\nu/kT\right)-1\right]^{-1}$. The brightness temperature
is defined as $T_{\rm B} = \left(c^{2}/2\nu^{2}k\right) I_{\nu}$ and is
directly proportional to the specific intensity $I_{\nu}$. The
general solution to the equation of transfer, assuming that
optical depth increases away from the observer is
\begin{equation}
\label{xfer}
T_{\rm B} =  T_{i} e^{-\tau_{0}} + \int_{0}^{\tau_{0}} J(T) e^{-\tau}
d\tau,
\end{equation}
where $T$ is the excitation temperature of a region which varies over the
optical depth interval from 0 to $\tau_{0}$, and $T_{i}$ is the incident
specific intensity of radiation on that region at $\tau_{0}$ in units of
brightness temperature. This equation can easily be integrated over regions of
constant excitation temperature. We assume regions of constant excitation
temperature along the line of sight in our first model.

\begin{figure}
\begin{center}
\resizebox{5in}{!}{\includegraphics{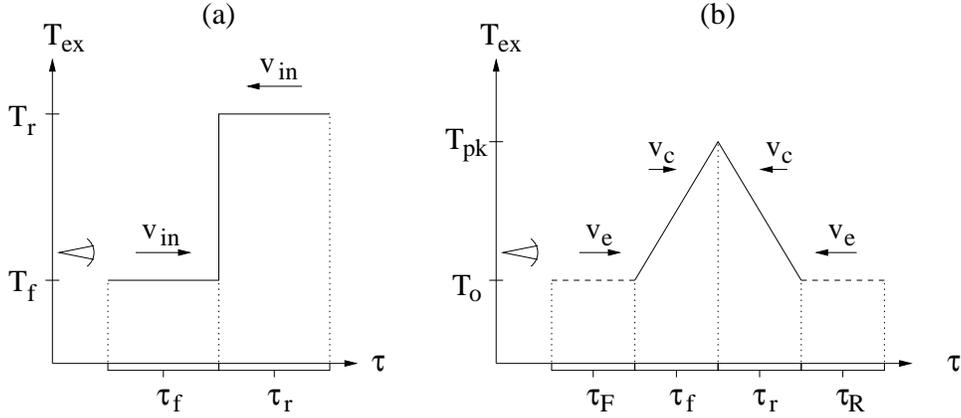}}
\end{center}
\caption{\label{models}Graphical representation of the analytic radiative
transfer models utilized in this paper. Each graph indicates the excitation
temperature for a line transition as a function of optical depth at some
frequency. Graph (a) is the two-layer model, which has a constant excitation
temperature $T_{f}$ for an optical depth of $\tau_{f}$ and a higher excitation
temperature $T_{r}$ for an optical depth of $\tau_{r}$. Each region of constant
excitation is traveling at velocity $v_{\rm in}$ toward the other region. The
symbol in the left indicates the position of the observer and the arrows
indicate the
direction of motion relative to the observer of each region. Graph
(b) is the hill model, which has an optional envelope of constant excitation
temperature
$T_{o}$ with an optical depth of $\tau_{F}$ on the side close to the observer,
and optical depth $\tau_{R}$ on the side opposite the observer (shown in dashed
lines). In the core of
the hill model the $J(T)$ rises linearly from $J(T_{0})$ to
$J(T_{\rm pk})$ over an optical depth of $\tau_{f}$ and then falls again to
$J(T_{0})$ over an optical depth of $\tau_{f}$. In the Rayleigh-Jeans limit
$J(T)$ is equal to the excitation temperature and the excitation profile is as
depicted, however at higher frequencies the lines rising to $T_{\rm pk}$ would
have some curvature.}
\end{figure}

The first
model, first discussed in \citet{mmtww}, we call the ``two-layer''
model. It is perhaps unfortunate that this name has been given to this model
and has stuck, as it implies that this is a model of two plane-parallel layers
along the line of sight. This is not true. The model actually applies to a line
of sight in which two regions with differing excitation temperatures are moving
toward each other. We assume the near region has a lower
excitation temperature than the far region. We assume both regions have a
velocity dispersion for the observed molecule $\sigma$ and a total optical
depth of 
$\tau_{0}$ at line
center. The excitation temperature of the front region is $T_{f}$, while that
of 
the rear region is $T_{r}$, and the regions are approaching each other with a
speed of $2v_{\rm in}$. The optical depth of each region at a velocity $v$ is
\begin{eqnarray}
\tau_{f}(v) & = & \tau_{0}\exp\left[-(v-v_{\rm LSR}-v_{\rm in})^{2}/2\sigma^{2}\right], \\
\tau_{r}(v) & = & \tau_{0}\exp\left[-(v-v_{\rm LSR}+v_{\rm
in})^{2}/2\sigma^{2}\right],
\end{eqnarray}
assuming that the average line of sight velocity of both regions is $v_{\rm
LSR}$. Figure~\ref{models} provides a graphical representation of the
model. The brightness temperature of the spectral line, obtained from the
equation of radiative transfer, is
\begin{equation}
\Delta T_{\rm B}(v) =  J(T_{f})\left[1-e^{-\tau_{f}(v)}\right] +
J(T_{r})\left[1-e^{-\tau_{r}(v)}\right]e^{-\tau_{f}(v)} -
J(T_{b})\left[1-e^{-\tau_{r}(v)-\tau_{f}(v)}\right],
\end{equation}
where
$T_{b}$ is the background temperature. The parameters $\tau_{0}$, $\sigma$,
$T_{f}$, $T_{r}$, $v_{\rm LSR}$, and $v_{\rm in}$ are free parameters which can
be adjusted to fit blue-asymmetric line profiles. \citet{lmt2} has used this
analytic model to derive infall velocity estimates for 29 starless cores. They
find that infall velocities are typically on the order of a tenth of a
kilometer per second, comparable to the velocity
dispersions measured in these sources.

Equation~\ref{xfer} is also analytically integrable if $J(T)$
is a linear function of the optical depth $\tau$. Assuming a simple
linear function $J(T) = J_{1} +
\left[\left(J_{2}-J_{1}\right)/\tau_{0}\right]\tau$ we can
integrate the equation of transfer to obtain
\begin{equation}
T_{\rm B} = T_{i} e^{-\tau_{0}} + (J_{2}-J_{1})\frac{1-e^{-\tau_{0}}}{\tau_{0}}
+ J_{1} - J_{2}e^{-\tau_{0}}.
\end{equation}
In the above equation $\tau$ is a function of the Doppler velocity, so in order
to calculate the intensity of emission across a line profile $J(T)$ must be a
linear function of $\tau$ at every frequency.
It is possible to construct a series of regions
along the line of sight where $J(T)$ varies linearly with $\tau$ within each
region and analytically calculate the brightness of radiation at a given
frequency emitted from that line of sight using the above equation. We have
constructed our next model using the above technique and equations which we
feel approximates the excitation conditions seen along the line of sight in
infalling clouds.

\begin{figure}
\begin{center}
\resizebox{5in}{!}{\includegraphics{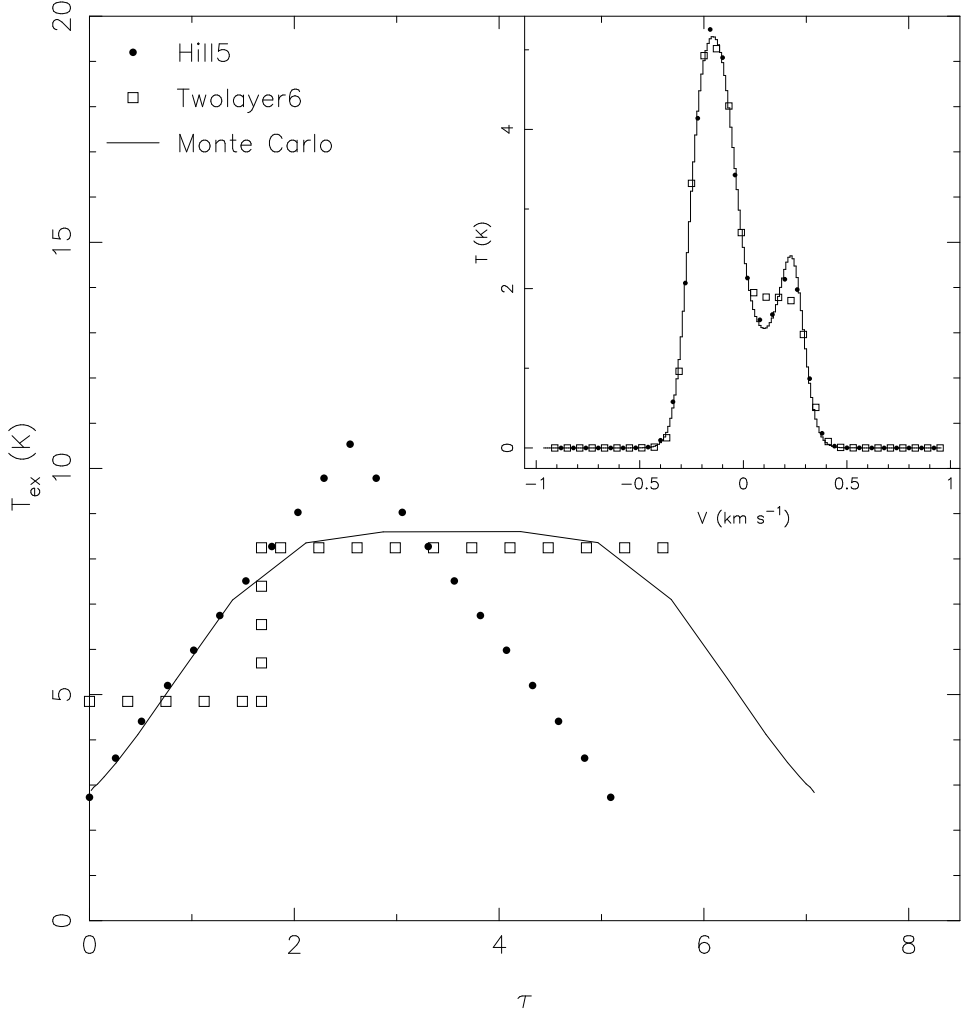}}
\end{center}
\caption{\label{excivtau}Line excitation temperature versus line center optical
depth for three radiative transfer models. The solid line indicates the
excitation temperature as
a function of optical depth for the \hcop~(\jequals{1}{0}) transition at line
center in a
starless cloud core whose density follows the expression $n(r) =
n_{0}/\left[1+\left(r/r_{0}\right)^{\alpha}\right]$, with $\alpha$ of 4, and $r_{0}$
of $5\times 10^{16}$~cm. The cloud has an velocity
dispersion of 0.1~km~s$^{-1}$
as well as a constant infall at all radii of 0.1~km~s$^{-1}$. The
\hcop~(\jequals{1}{0}) line profile
generated using the \citet{hv} radiative transfer model is shown in the upper
right corner of the figure. The filled circles
indicate the excitation profile of the best {\sc hill5} fit to the
emerging 
line profile of that cloud, while the open squares indicate the excitation
profile of the best {\sc twolayer6} model fit to the emerging line profile of
this simulated starless core. The line profiles generated by these models are
shown in the upper right hand corner of the figure.}
\end{figure}

The second model, which we call the ``hill'' model, is introduced for the first
time in this paper. 
This model consists of a core with a peak excitation temperature $T_{P}$ at the
center and an excitation temperature of $T_{0}$ at the near and far edges of
the core. The $J(T)$ drops linearly from $J(T_{P})$ at the center to $J(T_{0})$ at edges of the core,
forming a hill in the $J(T)$ profile. The
optical depth of the core is $\tau_{\rm C}$, and its infall
velocity is $v_{\rm C}$, while the
systematic velocity of the system is $v_{\rm LSR}$. A schematic representation
of this model is shown in figure~\ref{models}. We introduce this new type of
model because we believe it may be more analogous to the excitation profile we
are likely to be observing in starless cores. In figure~\ref{excivtau} we
compare the excitation profile as a function of optical depth at line center
for a simulated cloud, based on a Monte Carlo radiative transfer model, to the
best fit hill and two-layer models of the line
profile derived from that cloud. The excitation profile of our modeled starless
core has a significant slope, which the hill model can replicate, but the
two-layer model, due to its use of constant excitation temperature zones, can
not replicate. This improvement provides motivation for introducing a new
analytic radiative transfer model of infall.

In order to solve the
equation of radiative transfer we separate the hill model profile into two
regions along the line of sight: (1) The portion of the cloud in which the
excitation temperature is rising along the line of sight, whose optical depth
is $\tau_{f}$; And (2) The portion of the cloud in which the excitation
temperature is
falling along the line of sight, whose optical depth is $\tau_{r}$. The
optical depth as a function of line of sight
velocity for each of these regions is
\begin{eqnarray}
\tau_{f}(v) & = & \tau_{\rm C}\exp\left[-(v-v_{\rm LSR}-v_{\rm C})^{2}/2\sigma^{2}\right], \\
\tau_{r}(v) & = & \tau_{\rm C}\exp\left[-(v-v_{\rm LSR}+v_{\rm C})^{2}/2\sigma^{2}\right],
\end{eqnarray}
assuming a velocity dispersion of $\sigma$ in the entire cloud. We can now
solve the equation of transfer by integrating along the line of sight through
each of these two regions to derive the brightness temperature 
\begin{eqnarray}
\Delta T_{\rm B}(v) & = &
\left(J(T_{P})-J(T_{0})\right)\left[\left(1-e^{-\tau_{f}(v)}\right)/\tau_{f}(v)-
e^{-\tau_{f}(v)}\left(1-e^{-\tau_{r}(v)}\right)/\tau_{r}(v)\right]
\nonumber \\
 & & \mbox{ } + \left(J(T_{0})-J(T_{b})\right)\left[1-e^{-\tau_{r}(v)-\tau_{f}(v)}\right],
\end{eqnarray}
where $T_{b}$ is the background temperature. We believe the excitation
temperature gradient in this model might more closely approximate the
excitation profiles in observed clouds. The six free parameters in the
hill model are $\tau_{\rm C}$, $\sigma$, $T_{0}$, $T_{P}$,
$v_{\rm LSR}$, and $v_{\rm C}$. It is also possible to add an envelope of
constant excitation temperature moving with its own infall velocity. Assuming
this envelope has an excitation temperature of $T_{0}$, the optical depth of
the envelope and the velocity of the envelope adds two more free parameters,
creating a maximum of eight free parameters possible in the hill models.

In order to reduce the total number of free parameters we consider several
variants of each model by holding certain parameters fixed. We considered two
variants of the two-layer model, which we call {\sc twolayer5} and {\sc
twolayer6}, as well as 4 variants of the hill model, which we call {\sc hill5},
{\sc hill6}, {\sc hill6core}, and {\sc hill7}. The number is each variant name
refers to the number of free parameters in that variant. The parameterization
of the two-layer model variants are shown in table~\ref{twolayervar}, while
those of the hill model variants are shown in table~\ref{hillvar}.

\begin{deluxetable}{lllllll}
\tablecaption{\label{twolayervar}Variants of the two-layer model}
\tablecolumns{7}
\tablewidth{0pt}
\tablehead{\colhead{Variant Name} & \colhead{$\tau_{0}$} & \colhead{$v_{\rm
LSR}$} & \colhead{$v_{\rm in}$} & \colhead{$\sigma$} & \colhead{$T_{f}$} &
\colhead{$T_{r}$}}
\startdata
{\sc twolayer5} & Free & Free & Free & Free & $T_{b}$ & Free \\
{\sc twolayer6} & Free & Free & Free & Free & Free & Free \\
\enddata
\tablecomments{$\tau_{0}$ is the line center optical depth, $v_{\rm LSR}$ is
the line of sight velocity of the system, $v_{\rm in}$ is the infall velocity
of the system, $\sigma$ is the velocity dispersion of the molecular species,
$T_{f}$ is the excitation temperature of the front layer, and $T_{r}$ is the
excitation temperature of the rear layer.}
\end{deluxetable}

\begin{deluxetable}{lllllllll}
\tablecaption{\label{hillvar}Variants of the hill model}
\tablecolumns{9}
\tablewidth{0pt}
\tablehead{\colhead{Varient Name} & \colhead{$\tau_{\rm E}$} &
\colhead{$\tau_{\rm C}$} & \colhead{$v_{\rm LSR}$} &
\colhead{$v_{\rm E}$} & \colhead{$v_{\rm C}$} & \colhead{$\sigma$} &
\colhead{$T_{0}$} & \colhead{$T_{\rm pk}$}}
\startdata
{\sc hill5} & 0 & Free & Free & 0 & Free & Free & $T_{b}$ & Free \\
{\sc hill6} & Free & Free & Free & Free & 0 & Free & $T_{b}$ & Free \\
{\sc hill6core} & Free & Free & Free & 0 & Free & Free & $T_{b}$ & Free \\
{\sc hill7} & Free & Free & Free & Free & 0 & Free & Free & Free \\
\enddata
\tablecomments{$\tau_{\rm E}$ is the line center optical depth of the optional
envelope, $\tau_{\rm C}$ is the line center optical depth of the cloud, $v_{\rm
LSR}$ is the line of sight velocity of the system, $v_{\rm E}$ is the infall
velocity of the envelope, $v_{\rm C}$ is the infall velocity of the cloud,
$\sigma$ is the velocity dispersion of the molecular species, $T_{0}$ is the
excitation temperature of the edge of the cloud as well as the optional
envelope, and $T_{\rm pk}$ is the peak excitation temperature in the cloud.}
\end{deluxetable}

\section{Physical Models}
\label{physicalmodels}
The analytic radiative transfer models do an excellent job of fitting
blue-asymmetric line profiles \citep{lmt2}. Their efficacy in determining
physical parameters of the
molecular cloud, such as infall velocity, has not yet been
studied. In this paper we study how effectively we can recover the physical
parameters of modeled clouds using both the two-layer and hill analytic
radiative transfer models. We then use this information to determine the point
at which the systematic errors of the model dominate over random observational
errors. 

In our first three simulations, we modeled molecular clouds with a density
law of the form 
\begin{equation}
n(r) = \frac{n_{0}}{1+\left(r/r_{0}\right)^{\alpha}},
\end{equation}
where $n_{0}$ is the central density, $r_{0}$ is the turnover radius, and
$\alpha$ is the power law of the density profile at large radii
\citep{tmcwc}. This model reproduces the flat density region (at $r<r_{0}$) and
power law region (at $r>r_{0}$) required to model starless cores. It also is in
good agreement with the density profile of a Bonnor-Ebert sphere \citep{tmcw}
as well as the ``Plummer-like'' density model of \citet{ww}. The clouds
modeled in this paper have peak densities ($n_{0}$) of $6\times
10^{4}$~cm$^{-3}$, a turnover radius ($r_{0}$) of $5\times 10^{16}$~cm, an
outer radius of $2\times 10^{17}$~cm, and an $\alpha$ 
of~4. We chose to model \hcop\ rotational line emission from these clouds, as
it is often
moderately optically thick in starless cores. We assumed that the kinetic
temperature of the gas is 10~K throughout the cloud. We set
the relative abundance of
\hcop\ at $2\times 10^{-9}$ \citep{dbjg} and the \hcop/\hthcop\ abundance
ratio to 64. Depletion is not included here,
however we do investigate the effect of depletion in
\S\ref{depletionbeam}.
There are four velocity
laws which we investigate in this paper. The first (Simulation A) is constant
infall velocity
for all radii in the cloud. The second (Simulation B) is a constant infall
velocity for $r>r_{0}$ and 0 velocity for $r<r_{0}$. The third (Simulation C)
is constant
infall velocity for $r<r_{0}$ and 0 velocity for $r>r_{0}$. These simulations
are summarized in Table~\ref{sims}. These
velocity laws were chosen because they are simple velocity laws to
investigate. Our analytic radiative transfer models define a characteristic
single infall velocity which is analogous to Simulations A, B, and C. We set
the velocity dispersion to 0.1~km~s$^{-1}$ over the
entire cloud and typically allow the velocities within the cloud to be up to
twice the velocity dispersion. These results should all scale with the velocity
dispersion and the peak line brightness.
In addition to the simulations described above, we also simulated a different
density profile based on the best fit Bonnor-Ebert model of B68
\citep{all2}. The parameters of this simulation (Simulation~D) are discussed in
\S\ref{besphere}.

\begin{deluxetable}{llll}
\tablecaption{\label{sims}Density and velocity laws of the simulated clouds}
\tablecolumns{4}
\tablewidth{0pt}
\tablehead{\colhead{Simulation} & \colhead{Density Law} & \colhead{Velocity Law} & \colhead{Values}}
\startdata
A & TMCWC & $v_{\rm in} = a$ & $a =$
0.00, 0.02, \ldots, 0.2 km~s$^{-1}$ \\
B & TMCWC & $v_{\rm in} = \left\{ \begin{array}{ll} 0 & \mbox{if $r<r_{0}$} \\ a &
\mbox{if $r\geq r_{0}$} \end{array} \right. $ & $a =$
0.00, 0.02, \ldots, 0.2 km~s$^{-1}$ \\
C & TMCWC & $v_{\rm in} = \left\{ \begin{array}{ll} a & \mbox{if $r<r_{0}$} \\ 0 &
\mbox{if $r\geq r_{0}$} \end{array} \right. $ & $a =$
0.00, 0.02, \ldots, 0.2 km~s$^{-1}$ \\
D & BE & $v_{\rm in} = a$ & $a =$ 0.000, 0.024, \ldots, 0.24 km~s$^{-1}$  \\
\enddata
\tablecomments{The density laws are described in detail in
\S\ref{physicalmodels}. TMCWC refers to the flattened power law profile
developed in \citet{tmcwc}. BE refers to the exact solution of a hydrostatic
isothermal pressure-confined self-gravitating sphere, otherwise known as a
Bonnor-Ebert sphere.}
\end{deluxetable}

We use these simulated starless cloud cores to synthesize
\hcop~(\jequals{1}{0}) 
and \hcop~(\jequals{3}{2}) molecular line emission using the \citet{hv} 1-D
Monte Carlo radiative transfer model. The simulated clouds were divided into 20
spherical shells, each of uniform density, kinetic temperature, radial
velocity,
and with a radial size of $10^{16}$~cm. The level populations
within the first 21 rotational levels of the \hcop\ molecule were calculated
using the Einstein-A and \hcop\ with H$_{2}$ collisional rate coefficients from
\citet{monteiro} and \citet{green}. We then integrate along the central line of
sight to find the  \jequals{1}{0}\ and
\jequals{3}{2}\ molecular line spectral profiles of \hcop\ because they are
frequently observed, and
often have blue-asymmetric line profiles attributed to infall
\citep{gezc,nw,nwb,nmwb,brc1}. We then fit these synthesized spectra with the
two-layer and hill analytic radiative transfer models to check the ability of
the analytic models to recover physical parameters within the modeled cloud.

Several algorithms exist for minimizing multidimensional functions. There is
always the danger of minimizing to a ``local'' minimum (meaning the lowest
value within some finite region of the function to be minimized) instead of the
``global'' minimum (meaning the lowest function value over the entire parameter
space) and it becomes more difficult to avoid local minima as your function
goes to higher dimensions. There are several strategies to avoid local minima,
and we have chosen a hybrid minimization algorithm to do just that.
We use the Differential Evolution
(DE) algorithm of \citet{devo} initially to separate the local minima from the
global minimum, then we use the
Nelder-Mead simplex method \citep{nm} to optimize the fit. The DE minimization
algorithm is applicable to many astronomical problems, and is similar in scope
to simulated annealing. DE is an evolutionary algorithm in that several
parameter sets are randomly selected and these are randomly modified over
several ``generations'' until acceptable convergence has been achieved. The
variability in each parameter is scaled by the variation of the solution sets
in each generation, so as the algorithm converges the region of the search
decreases without the need to include a decreasing ``temperature'' as in
simulated annealing. DE is also
superior to simulated annealing in that it is more likely to achieve
convergence, is a simpler algorithm, and has fewer adjustable parameters
\citep{devo}. 

\section{Results}
We made four simulations of infalling clouds and used the \citet{hv}
Monte Carlo radiative transfer model to extract \hcop~(\jequals{1}{0}) and
\hcop~(\jequals{3}{2}) line profiles from those simulations. In this section we
will discuss how effective our analytic models are at deriving the input
velocities in those simulation, quantify the relative contributions of both
random and systematic errors in model fits of simulated data sets, and apply
this model to an observed infall profile in a source which has been rigorously
modeled using Monte Carlo radiative transfer techniques.

\subsection{Simulation A --- Constant Infall}
\begin{figure}
\begin{center}
\resizebox{5in}{!}{\includegraphics{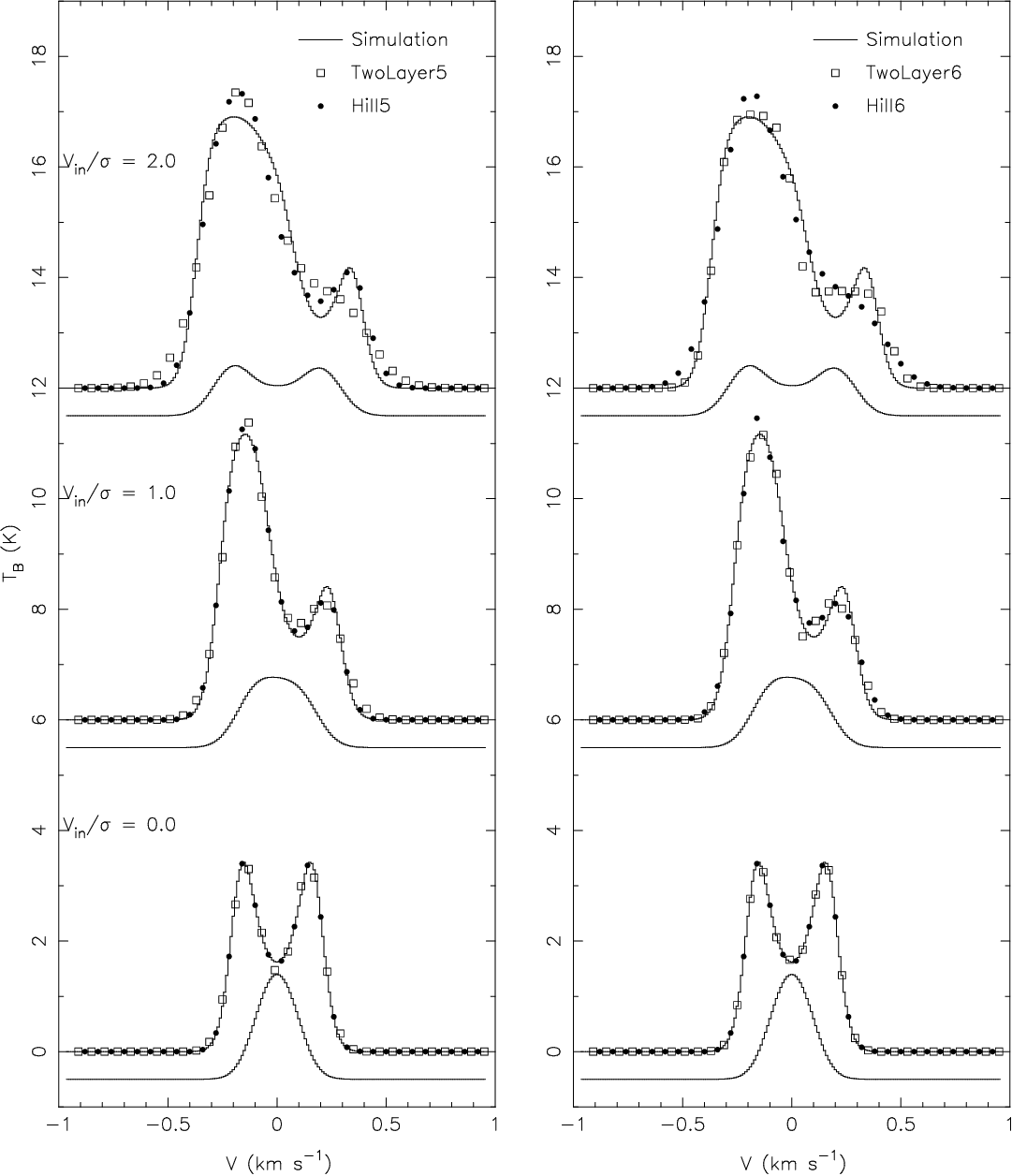}}
\end{center}
\caption{\label{majo} Monte Carlo simulated spectra (solid lines) of the
\hcop~(\jequals{1}{0}) and \hthcop~(\jequals{1}{0}) (multiplied by a factor of
5) emission from
Simulation~A (Constant Infall). The infall velocity in the lowest set of
spectra is
0~km~s$^{-1}$, those in the middle are of a cloud infalling at 0.1~km~s$^{-1}$,
while those at the top are of a cloud infalling at 0.2~km~s$^{-1}$. The
analytic model fits are overlaid on each \hcop\ spectrum. The open squares on
the left indicate the {\sc twolayer5} fits, while the filled circles are the
{\sc hill5} fits. On the right the open squares indicate the {\sc twolayer6}
fits, while the filled circles indicate the {\sc hill6} fit.}
\end{figure}

\begin{figure}
\begin{center}
\resizebox{5in}{!}{\includegraphics{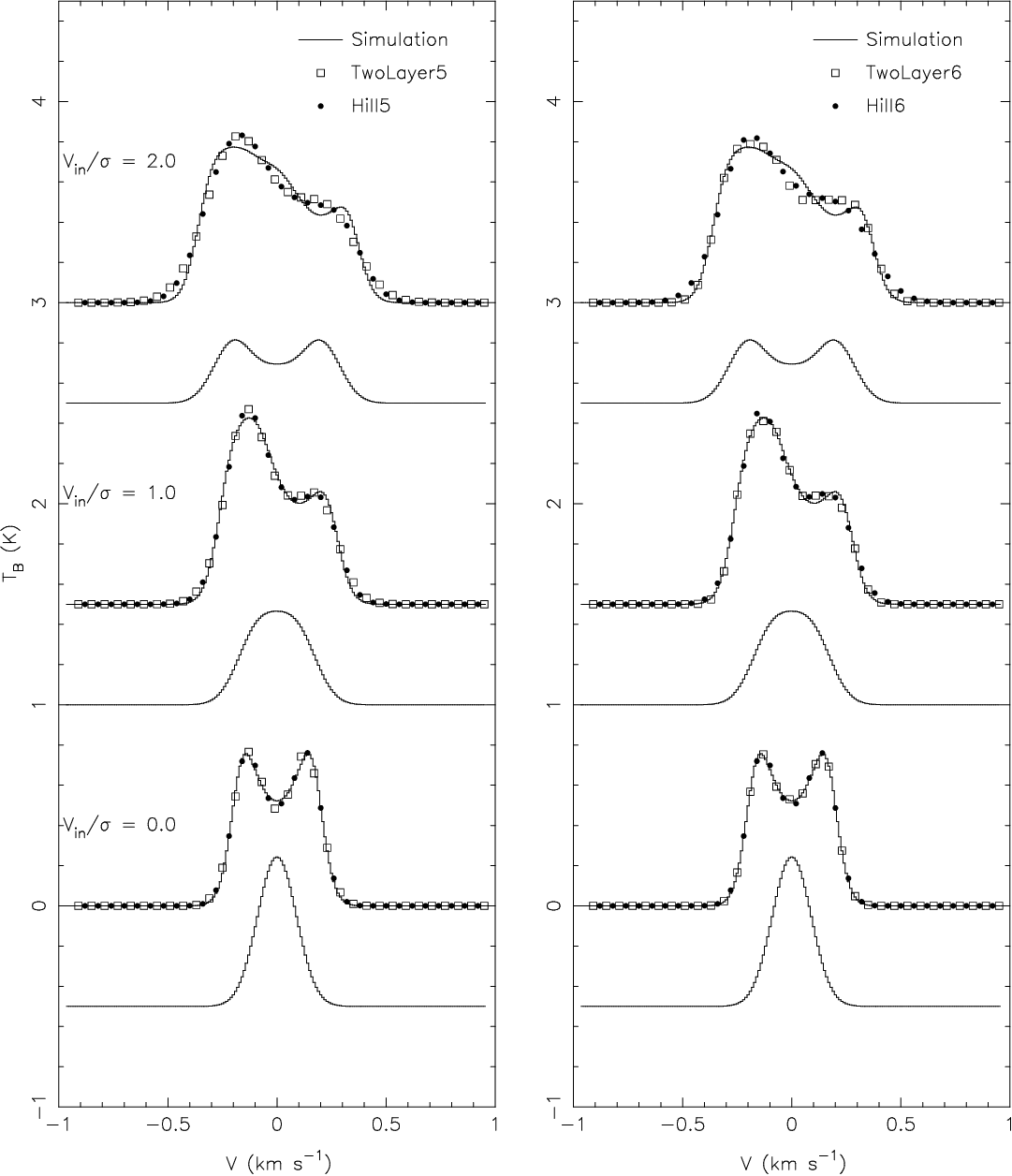}}
\end{center}
\caption{\label{majt} Monte Carlo simulated spectra (solid lines) of the
\hcop~(\jequals{3}{2}) and \hthcop~(\jequals{3}{2}) (multiplied by a factor of
40) emission from
Simulation~A (Constant Infall). The infall velocity in the lowest set of
spectra is
0~km~s$^{-1}$, those in the middle are of a cloud infalling at 0.1~km~s$^{-1}$,
while those at the top are of a cloud infalling at 0.2~km~s$^{-1}$. The
analytic model fits are overlaid on each \hcop\ spectrum. The open squares on
the left indicate the {\sc twolayer5} fits, while the filled circles are the
{\sc hill5} fits. One the right the open squares indicate the {\sc twolayer6}
fits, while the filled circles indicate the {\sc hill6} fit.}
\end{figure}

Simulation A, which has a constant infall velocity for all cloud radii,
has asymmetric line profile typical of those seen in infalling clouds. The
\hcop~(\jequals{1}{0}) line profiles are shown in figure~\ref{majo}, along with
their optically thin \hthcop~(\jequals{1}{0}) counterparts. The spectra in the
bottom on the left and right side are for a simulation with no infall. The
middle spectra on the left and right are of a simulation with an infall
velocity of 0.1~km~s$^{-1}$. The top spectra on the left and right are from a
simulation with an infall velocity of 0.2~km~s$^{-1}$. Overlain on the
\hcop~(\jequals{1}{0}) spectra on the left panel are the {\sc twolayer5} and
{\sc hill5} best fit models, while on the right the {\sc twolayer6} and {\sc
hill6} spectra are shown. Figure~\ref{majt} depicts \hcop~(\jequals{3}{2}) and
\hthcop~(\jequals{3}{2}) spectra from the same simulations along with their
{\sc twolayer5}, {\sc hill5}, {\sc twolayer6}, and {\sc hill6} best fit spectral
line profiles. Two variants are not shown in these fits, the {\sc
hill6core} model best fit in most cases has a $\tau_{\rm E}$ which is very
close to 0, yielding the same parameters and spectral line shape as the {\sc
hill5} model. The {\sc hill7} model always manages a very good fit to the
spectral line shape, however in some cases, especially in spectra with a small
red component that is not well separated from the blue component (such as the
middle and top spectra in figure~\ref{majt}), the parameters of the {\sc hill7}
model are not very well constrained. A wide range of infall velocities, optical
depths, and excitation temperatures in the seven parameter {\sc hill7} model
can yield the same line profile.

There are several interesting features in these spectral line fits. When the
infall velocity is greater than the velocity dispersion the \hthcop\ line
profiles become double peaked. This is not due to self-absorption as the lines
are optically thin. It is due to the fact that the molecular gas along the
central line of sight is traveling at two velocities separated by more than the
velocity dispersion, resulting in two peaks in the optically thin
component. This feature is unlikely to be observed in
\hthcop~(\jequals{3}{2}) lines as they are too weak to be easily detected. Also
interesting is that while the
dip between the red and blue peaks of the spectral line profile is deep, both
the hill variants and the two-layer variants do a good job a reproducing the
spectral line, but as the red peak drops in intensity relative to the blue peak
the analytic model fits tend to produce a spectral line with a single blue peak
and a red shoulder with no local maximum but only a flattening of the
slope. The two-layer model variants tend to produce this type of spectrum at
lower infall speeds
than the hill model variants (Figure~\ref{majo}). We call the spectra
with distinct blue and red peaks ``dip'' spectra and those with only a
flattening of the slope, ``shoulder'' spectra. We also see that none of the
models can reproduce a spectral line which has a minimal red peak superimposed
on a broadly asymmetric line (as in the top spectra of
figure~\ref{majt}). These models tend to produce spectra that are similar in
shape to two overlapping gaussians of equal width. If we try to fit a spectral
line that differs greatly from one that can be obtained by adding two gaussians
of equal width, then we tend to not get good line fits with any of the analytic
models. 

\begin{figure}
\begin{center}
\resizebox{5in}{!}{\includegraphics{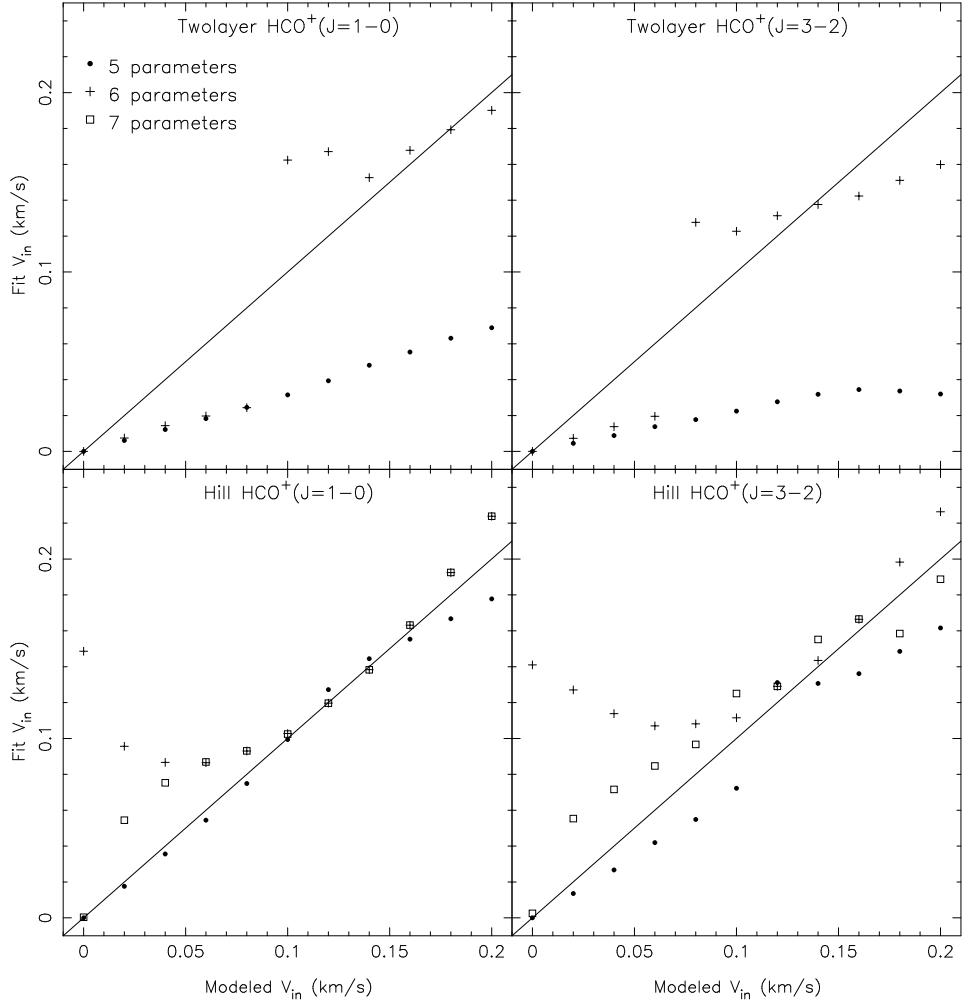}}
\end{center}
\caption{\label{maf} Infall velocity fits. The above figures indicate the
infall velocities obtained by fitting {\sc twolayer5}, {\sc twolayer6}, {\sc
hill5}, {\sc hill6}, and {\sc hill7} models to the Monte
Carlo simulated
spectra in Simulation~A (Constant Infall). The left hand panels are the results
of
fitting the \hcop~(\jequals{1}{0}) spectra, and the right hand panels are the
results of fitting the \hcop~(\jequals{3}{2}) spectra. The top two figures
indicate the infall velocities obtained from the two-layer models. Below the
two-layer results are the results of the hill models. The symbols indicate the
number of free parameters, closed circles indicate the 5 free parameter models
({\sc twolayer5} and {\sc hill5}). Crosses indicate the 6 free parameter models
({\sc twolayer6} and {\sc hill6}). Open squares indicate the 7 free parameter
model ({\sc hill7}). The {\sc twolayer6} model often has two local minima, one
of which is usually fit by the {\sc twolayer5} model. We show the other local
minimum when it is available in this plot.}
\end{figure}

In figure~\ref{maf} we show the quality of the infall velocity fit attained by
the analytic model fits to the \hcop~(\jequals{1}{0}) line profiles in
this simulation (left side) as well as the model fits to the
\hcop~(\jequals{3}{2}) line profiles in this simulation (right side). In this
simulation, for the entire infall velocity range up to twice the velocity
dispersion, the {\sc hill5} model tends to reproduce the modeled infall
velocity with an RMS systematic error of less than 0.01~km~s$^{-1}$ in the
\hcop~(\jequals{1}{0}) lines and 0.02~km~s$^{-1}$ in the \hcop~(\jequals{3}{2})
lines. The
{\sc twolayer5} model tends to consistently underestimate the infall velocity
by over a factor of 2 in both transitions when it produces good line fits. The
{\sc twolayer6} model fits the infall velocity with an RMS error of
0.03~km~s$^{-1}$ in both the 
\hcop~(\jequals{1}{0}) lines and the
\hcop~(\jequals{3}{2}) lines. Generally
the {\sc twolayer6} model does a better
job in cases where the spectrum is dominated by the blue component and the
relative size or integrated intensity of the red component is small. This
happens to be the true for cases with higher infall velocities in the
\hcop~(\jequals{1}{0}) and \hcop~(\jequals{3}{2}) spectra in this
simulation. The {\sc twolayer6} model also performs well when the excitation
temperature of the rear layer matches the simulated excitation temperature in
the cloud. The {\sc hill6} model does a good
job at fitting the infall
velocity for infall velocities greater than the velocity dispersion, but tends
to greatly overestimate the infall velocity in cases where it is smaller than
the velocity dispersion. The {\sc hill7} model also tends to reproduce the
infall velocity fairly well, but not as well as the {\sc hill5} model.

It
should be noted that a good fit to the line profile does not always translate
to a good fit to the physical parameters, such as the infall velocity, in the
simulated clouds. In the case of the {\sc twolayer6} models, often the dip
solution is a good fit to the line profile, but a worse fit to the infall
velocity than the shoulder solution. The {\sc hill7} model is able to fit just
about any line profile, but often yields worse infall velocities than {\sc
hill5} model fits with larger minimized $\chi^{2}$.
In summary, the infall velocities for Simulation~A (constant infall) are most
accurately reproduced by the {\sc hill5} model with a RMS deviation of 0.016
km~s$^{-1}$ over 22 cases. The {\sc twolayer6} model is significantly worse
with an overall RMS deviation of 0.030~km~s$^{-1}$ over 22 cases. 

\subsection{Simulation B --- Envelope Infall}
\begin{figure}
\begin{center}
\resizebox{5in}{!}{\includegraphics{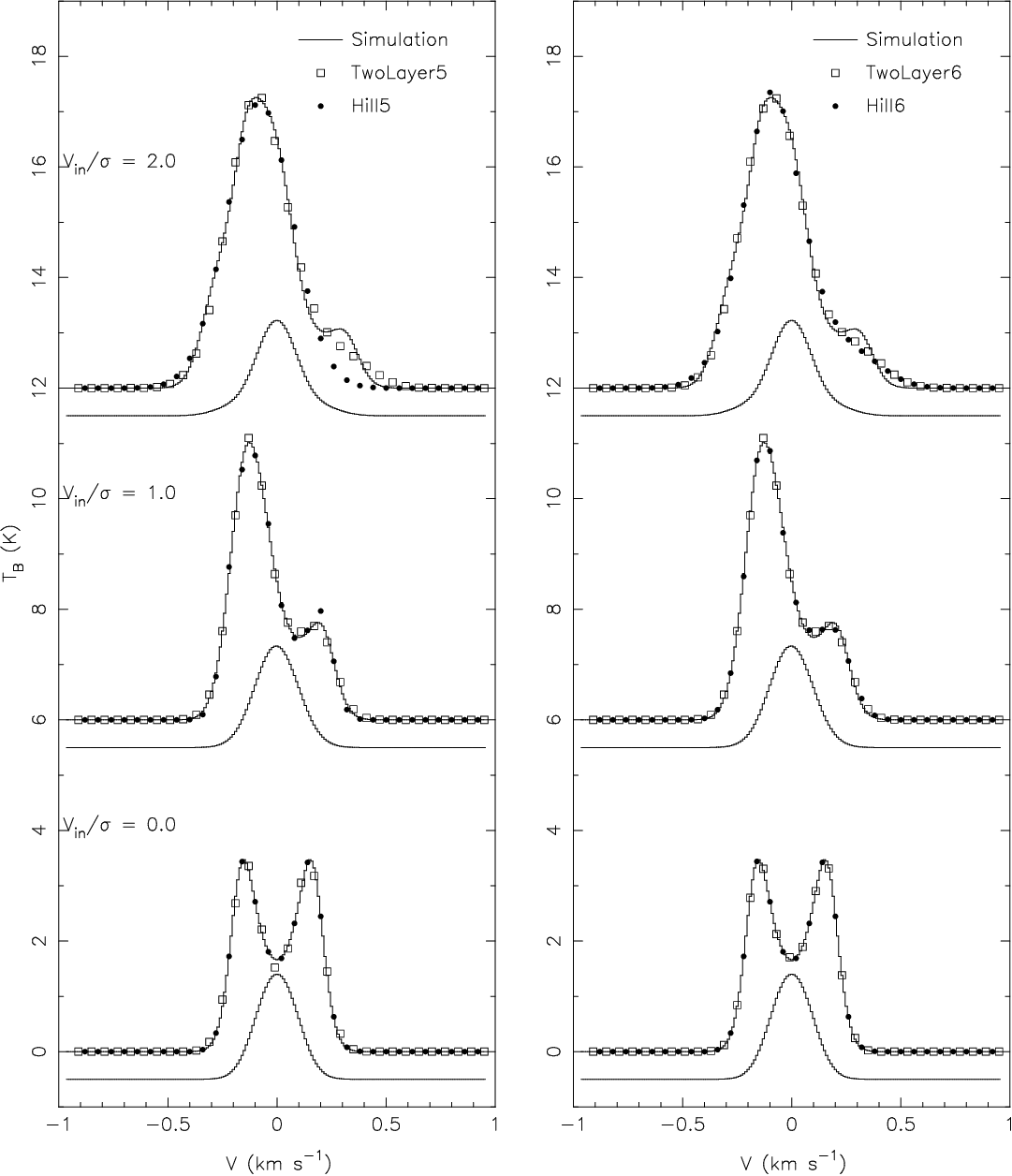}}
\end{center}
\caption{\label{mbjo} Monte Carlo simulated spectra (solid lines) of the
\hcop~(\jequals{1}{0}) and \hthcop~(\jequals{1}{0}) (multiplied by a factor of
5) emission from
Simulation~B (Envelope Infall). The infall velocity in the lowest set of
spectra is
0~km~s$^{-1}$, those in the middle are of a cloud infalling at 0.1~km~s$^{-1}$,
while those at the top are of a cloud infalling at 0.2~km~s$^{-1}$. The
analytic model fits are overlaid on each \hcop\ spectrum. The open squares on
the left indicate the {\sc twolayer5} fits, while the filled circles are the
{\sc hill5} fits. One the right the open squares indicate the {\sc twolayer6}
fits, while the filled circles indicate the {\sc hill6} fit.}
\end{figure}

\begin{figure}
\begin{center}
\resizebox{5in}{!}{\includegraphics{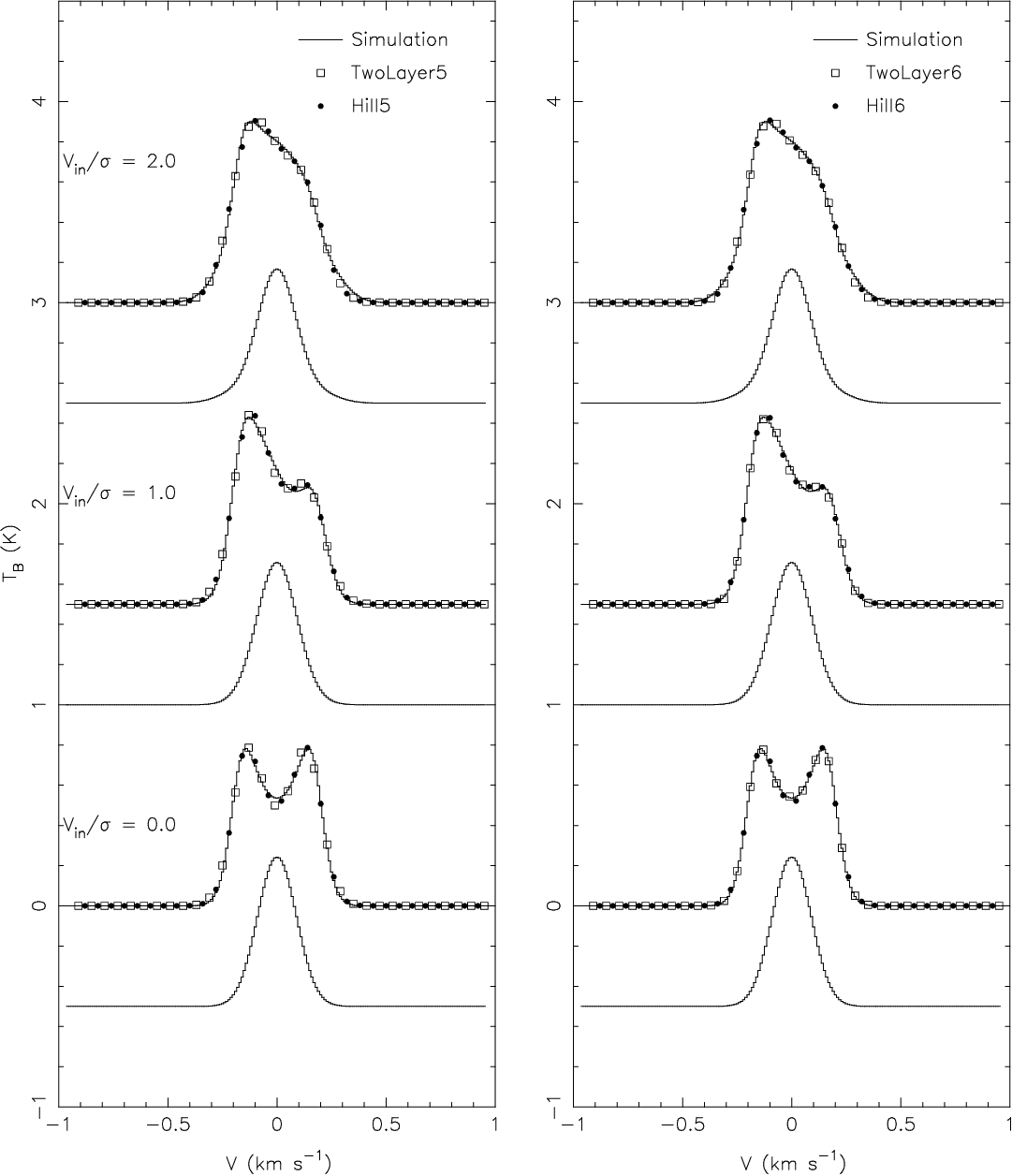}}
\end{center}
\caption{\label{mbjt} Monte Carlo simulated spectra (solid lines) of the
\hcop~(\jequals{3}{2}) and \hthcop~(\jequals{3}{2}) (multiplied by a factor of
40) emission from
Simulation~B (Envelope Infall). The infall velocity in the lowest set of
spectra is
0~km~s$^{-1}$, those in the middle are of a cloud infalling at 0.1~km~s$^{-1}$,
while those at the top are of a cloud infalling at 0.2~km~s$^{-1}$. The
analytic model fits are overlaid on each \hcop\ spectrum. The open squares on
the left indicate the {\sc twolayer5} fits, while the filled circles are the
{\sc hill5} fits. One the right the open squares indicate the {\sc twolayer6}
fits, while the filled circles indicate the {\sc hill6} fit.}
\end{figure}

Simulation~B, which has a constant infall velocity for $r>r_{0}$ and no
infall for $r<r_{0}$, has asymmetric line profiles typical of those seen in
infalling clouds. Figure~\ref{mbjo} shows the \hcop~(\jequals{1}{0}) line
profiles for three different infall velocities. The self-absorbed
solid line in the lowest left and right panels is the line profile for a cloud
with no
infall, as well as the optically thin
\hthcop~(\jequals{1}{0}) line profile. In the middle left and right are the
\hcop\ and \hthcop~(\jequals{1}{0}) line profiles for a cloud with an infall
velocity of 0.1~km~s$^{-1}$, equal to the velocity dispersion in that
cloud. The upper left and right panels are the \hcop\ and
\hthcop~(\jequals{1}{0}) line profiles for a cloud with an infall velocity of
0.2~km~s$^{-1}$. These line profiles are slightly narrower than those seen in
Simulation~A (Constant Infall). This is due to the fact that the cores, where
the most dense gas
is located, are stationary and not producing any broad line wings. As in our
previous simulation we have overlaid the {\sc twolayer5}, {\sc hill5}, {\sc
twolayer6}, and {\sc hill6} best fit to the \hcop\ spectra. Figure~\ref{mbjt}
indicates these same results for the \hcop~(\jequals{3}{2}) and
\hthcop~(\jequals{3}{2}) spectra in Simulation~B (Envelope Infall).

\begin{figure}
\begin{center}
\resizebox{5in}{!}{\includegraphics{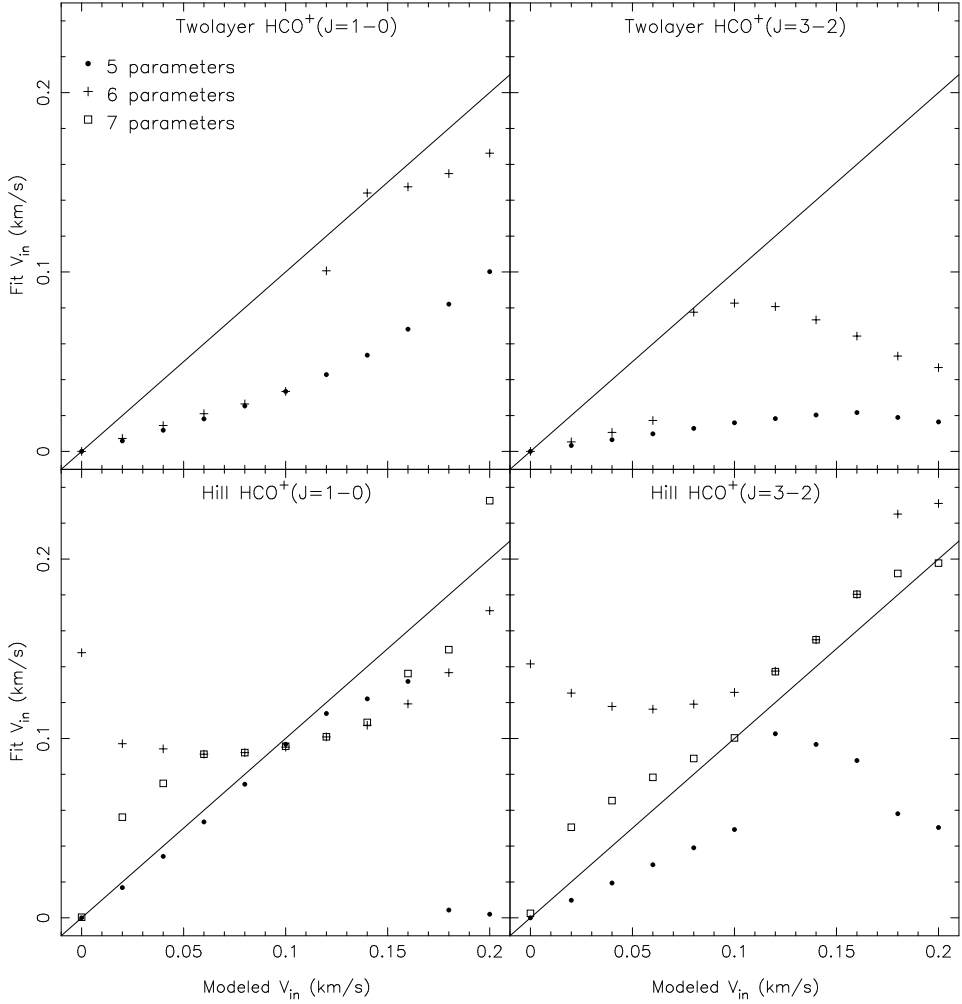}}
\end{center}
\caption{\label{mbf} Infall velocity fits. The above figures indicate the
infall velocities obtained by fitting {\sc twolayer5}, {\sc twolayer6}, {\sc
hill5}, {\sc hill6}, and {\sc hill7} models to the Monte
Carlo simulated
spectra in Simulation~B (Envelope Infall). The left hand panels are the results
of
fitting the \hcop~(\jequals{1}{0}) spectra, and the right hand panels are the
results of fitting the \hcop~(\jequals{3}{2}) spectra. The top two figures
indicate the infall velocities obtained from the two-layer models. Below the
two-layer results are the results of the hill models. The symbols indicate the
number of free parameters, closed circles indicate the 5 free parameter models
({\sc twolayer5} and {\sc hill5}). Crosses indicate the 6 free parameter models
({\sc twolayer6} and {\sc hill6}). Open squares indicate the 7 free parameter
model ({\sc hill7}). The {\sc twolayer6} model often has two local minima, one
of which is usually fit by the {\sc twolayer5} model. We show the other local
minimum when it is available in this plot.}
\end{figure}

We investigate the efficacy of our models in matching the infall velocity of
the Simulation~B (Envelope Infall) clouds in figure~\ref{mbf}. The {\sc
twolayer5} model once
again underestimates the infall velocity in all cases and in both transitions,
although it monotonically increases throughout the simulation. The {\sc
twolayer6} model tends to yield the same result as the {\sc twolayer5} model
over many of the infall velocities, but jumps to a higher infall velocity which
in many cases approximates the correct infall velocity (especially in the
\hcop~(\jequals{1}{0}) fits). We investigated these jumps in the {\sc
twolayer6} model further and found that there are two local minima in the
$\chi^{2}$ surface for many spectrum shapes. 

\begin{figure}
\begin{center}
\resizebox{5in}{!}{\includegraphics{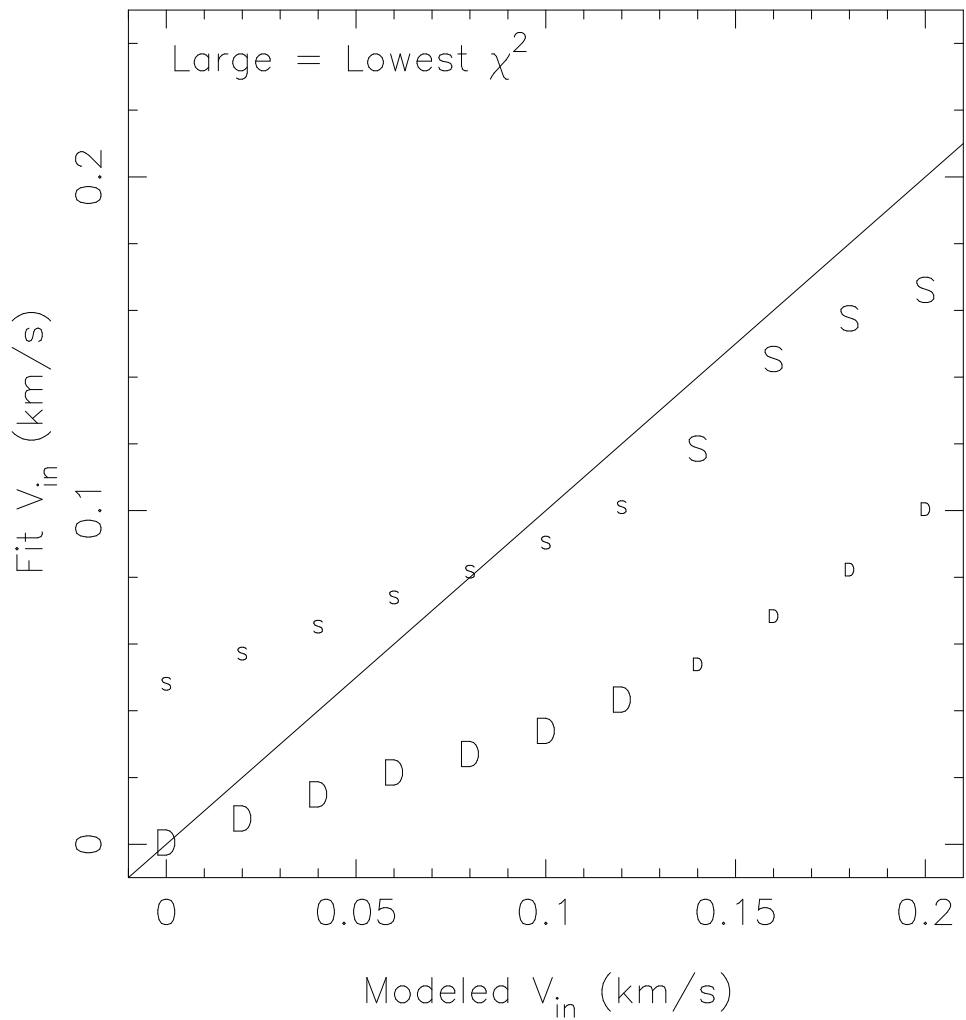}}
\end{center}
\caption{\label{dipshoulder} We find that there are often two local minima in
the $\chi^{2}$ plane of the {\sc twolayer6} fit to a spectrum. We trace each
local minimum for the \hcop~(\jequals{1}{0}) spectra in simulation~B here. The
``S'' points correspond to a local minimum that tends to produce a ``shoulder''
spectrum, while the ``D'' points corresponds to a local minimum that tends to
produce a ``dip'' spectrum. The larger letter indicates which minimum is the
global minimum for a particular infall velocity.}
\end{figure}

The {\sc twolayer6} minima we found
correspond to ``dip'' spectra, with two distinct peaks, and ``shoulder''
spectra with one peak and a flattening of the spectrum on the red side. These
two spectral line fits tend to have very different parameters for the infall
velocity which we show in figure~\ref{dipshoulder} for the
\hcop~(\jequals{1}{0}) lines generated in this simulation. The letter with the
larger size indicates the
fit with the lowest $\chi^{2}$, and the solid line indicates the expected value
if the infall velocity modeled analytically matches the infall velocity of the
cloud. In the \hcop~(\jequals{1}{0}) profiles, the ``dip'' solution is a better
fit to the line profile up to a modeled infall velocity of 0.12~km~s$^{-1}$,
however the ``shoulder'' solution yields a better fit to the infall velocity
for modeled infall velocities above 0.04~km~s$^{-1}$. A similar trend is seen
in the \hcop~(\jequals{3}{2}) results. The ``shoulder''
fits tend to have a higher infall velocity, and are often a much better match
to the infall velocity. The ``dip'' spectra are typically the equivalent to the
{\sc twolayer5} model results. The best ``dip'' spectrum fit of the {\sc
twolayer6} model usually results in an excitation temperature of the front
layer equal to the background temperature and an underestimate of the infall
velocity. We therefore recommend that one should attempt to find the local
minimum corresponding to a ``shoulder'' fit when using the {\sc twolayer6}
model, and that one can verify that one has found the ``shoulder'' fit by
making sure that the excitation temperature of the front layer is significantly
higher than the background temperature. 

Another trend revealed in
figure~\ref{mbf} is that the {\sc twolayer6} model does not
do a good job of fitting high velocity infall using the \hcop~(\jequals{3}{2})
line profile in this modeled cloud. We can see in the upper-most line profiles
of figure~\ref{mbjt} that the \hcop~(\jequals{3}{2}) line profiles do not
separate into multiple distinct components at high infall velocities in this
model. As the lines begin to blend into a profile that is not easily separable
into multiple components, the two-layer model begins to break down and does not
accurately match the actual infall velocity.

The lower panels of figure~\ref{mbf} show the hill model variants infall
velocities compared to the infall velocities in those same clouds. Once again
we see that {\sc hill5} does a good job of matching most of the infall
velocities using the \hcop~(\jequals{1}{0}) line profile, but tends to break
down at high velocities. Overall the {\sc hill5} model fits to the
\hcop~(\jequals{1}{0} match the simulated
infall velocities with an RMS error of 0.08~km~s$^{-1}$ over the full range of
simulated infall velocities, however for infall velocities less than 1.5 times
the velocity dispersion the RMS error is less than 0.01~km~s$^{-1}$. The {\sc
hill5} infall velocity fits break down at an even lower infall velocity in the
case of the
\hcop~(\jequals{3}{2}) line, however the overall RMS error in the infall
velocity fits for that transition are only 0.07~km~s$^{-1}$. We believe this is
because the contribution of the
red component is so low in these cases. When the magnitude of the red component
drops {\sc hill5} does not produce a good profile fit (see the topmost spectrum
of figure~\ref{mbjo}). We also believe that the {\sc hill5} model does a poor
job when the shoulder starts to disappear yielding a sloping asymmetric line
such as the uppermost spectrum of figure~\ref{mbjt}. Once again the {\sc hill6}
model does a poor job at fitting low infall velocities. The {\sc hill7} does a
good job at fitting all the infall velocities in both line transitions,
typically within 0.03~km~s$^{-1}$ of the modeled value. The error in the fit
begins to increase, especially in the case of the
\hcop~(\jequals{1}{0}) fit, as the infall velocity increases. This is due to
the diminishing magnitude of the red-shifted component. As the blue-shifted
component begins to dominate the {\sc hill7} model fit, the seven parameters of
the
model begin to overdetermine the line profile, resulting in a range of
solutions that fit equally well. In practice it is possible to avoid this by
reducing the number of parameters and using the {\sc hill6} model, however in
most observed infall
profiles the magnitudes of both the red and blue components are not as large as
seen in the top panel of figure~\ref{mbjo}\ \citep[cf.][]{lmt2}. In typical
line profiles, such as those seen in this spherically symmetric infalling
cloud, with a uniform infall velocity in the outer layers, the {\sc hill5} does
an excellent job at extracting the infall velocity, while the {\sc twolayer6}
does an adequate job for lines with a small red peak or narrow shoulder
component.

\subsection{Simulation C --- Core Infall}
\begin{figure}
\begin{center}
\resizebox{5in}{!}{\includegraphics{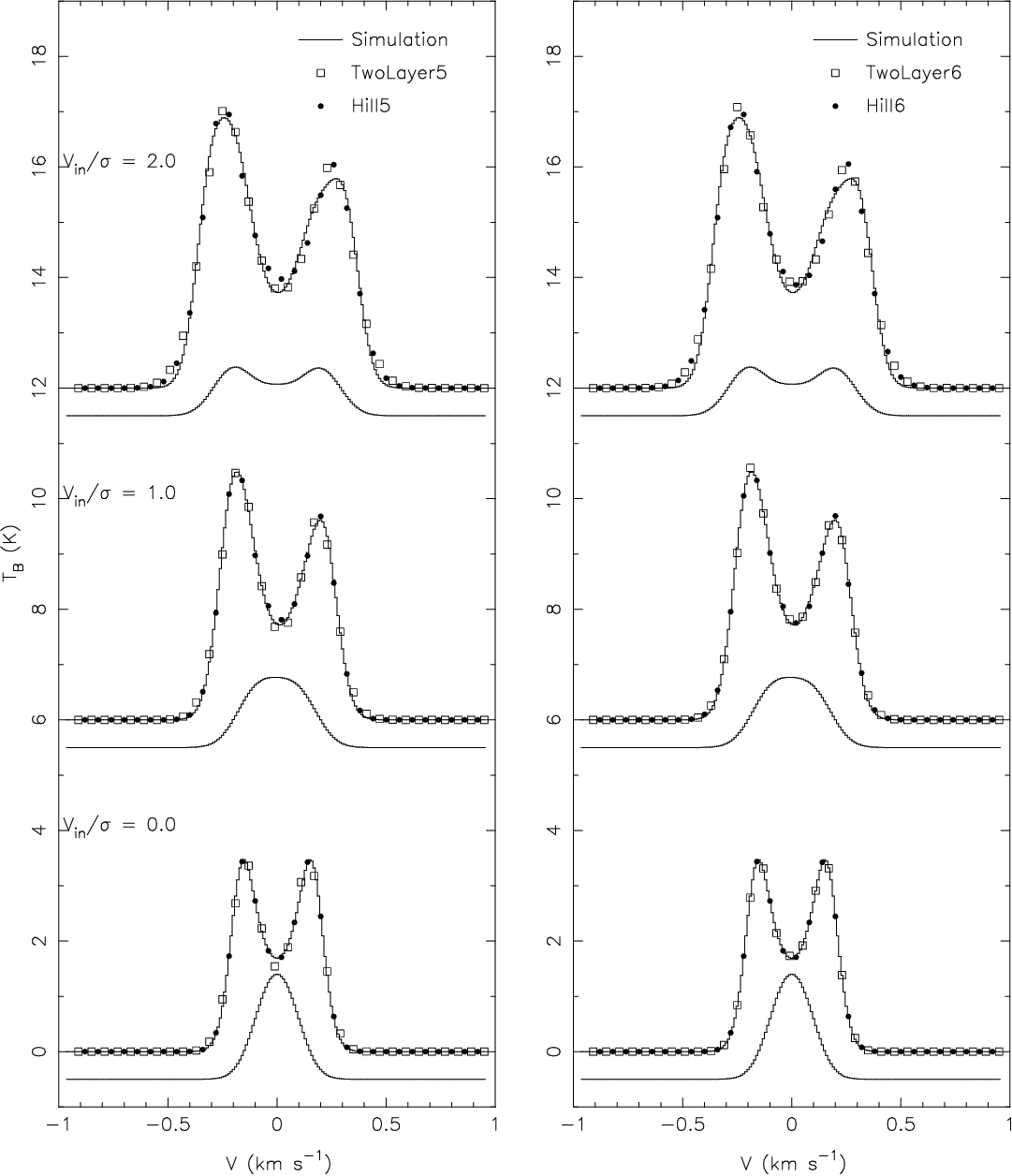}}
\end{center}
\caption{\label{mcjo} Monte Carlo simulated spectra (solid lines) of the
\hcop~(\jequals{1}{0}) and \hthcop~(\jequals{1}{0}) (multiplied by a factor of
5) emission from
Simulation~C (Core Infall). The infall velocity in the lowest set of spectra is
0~km~s$^{-1}$, those in the middle are of a cloud infalling at 0.1~km~s$^{-1}$,
while those at the top are of a cloud infalling at 0.2~km~s$^{-1}$. The
analytic model fits are overlaid on each \hcop\ spectrum. The open squares on
the left indicate the {\sc twolayer5} fits, while the filled circles are the
{\sc hill5} fits. One the right the open squares indicate the {\sc twolayer6}
fits, while the filled circles indicate the {\sc hill6} fit.}
\end{figure}

\begin{figure}
\begin{center}
\resizebox{5in}{!}{\includegraphics{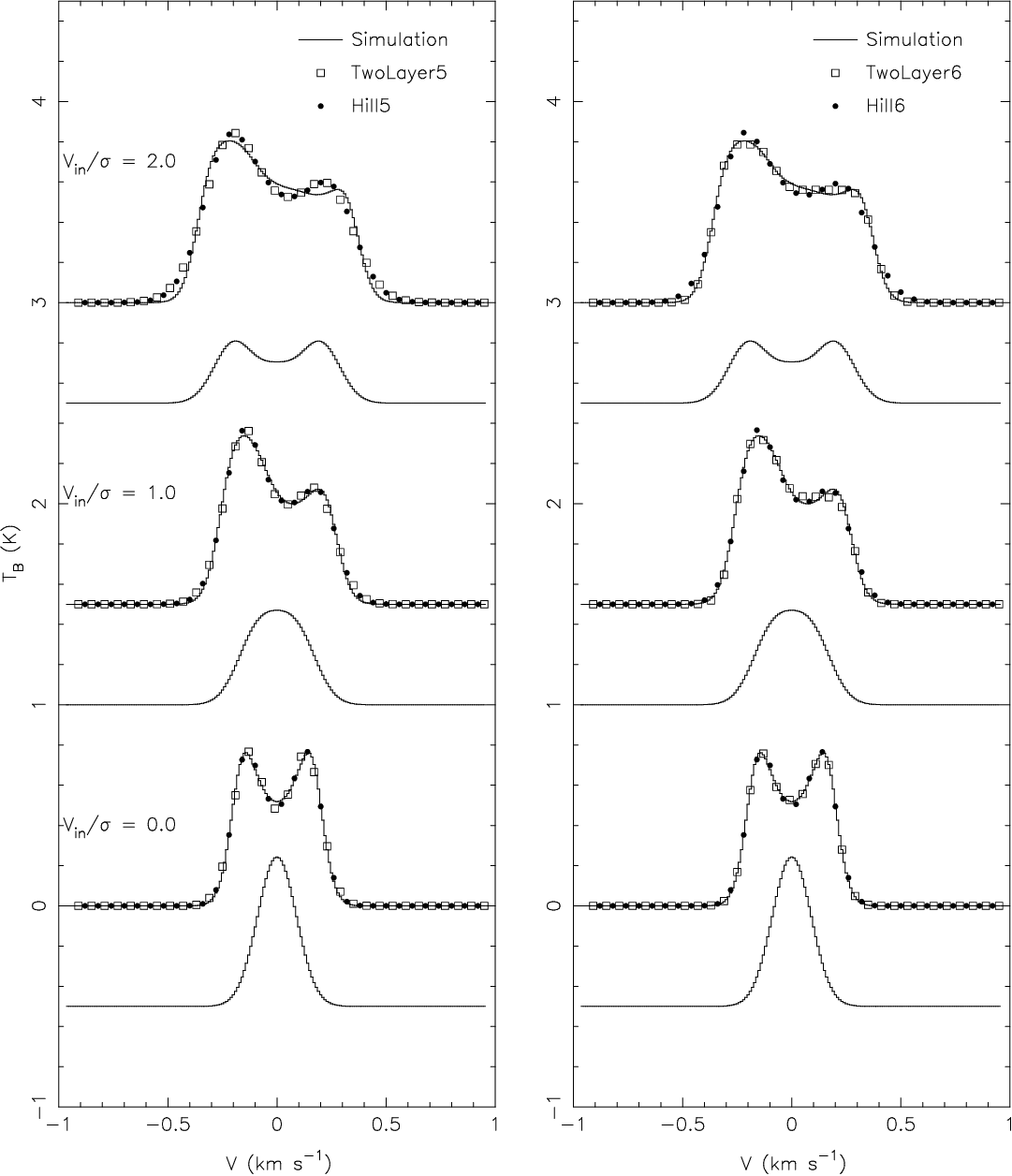}}
\end{center}
\caption{\label{mcjt} Monte Carlo simulated spectra (solid lines) of the
\hcop~(\jequals{3}{2}) and \hthcop~(\jequals{3}{2}) (multiplied by a factor of
40) emission from
Simulation~C (Core Infall). The infall velocity in the lowest set of spectra is
0~km~s$^{-1}$, those in the middle are of a cloud infalling at 0.1~km~s$^{-1}$,
while those at the top are of a cloud infalling at 0.2~km~s$^{-1}$. The
analytic model fits are overlaid on each \hcop\ spectrum. The open squares on
the left indicate the {\sc twolayer5} fits, while the filled circles are the
{\sc hill5} fits. One the right the open squares indicate the {\sc twolayer6}
fits, while the filled circles indicate the {\sc hill6} fit.}
\end{figure}

We confine infall to the core ($r<r_{0}$), and surround the infall region with
a static
envelope in Simulation~C. The resulting \hcop\ and
\hthcop~(\jequals{1}{0}) 
line 
profiles are shown in figure~\ref{mcjo}. The \hcop\ and
\hthcop~(\jequals{3}{2}) line
profiles are shown in figure~\ref{mcjt}. Once again modeled infall velocity,
this time in the core, is increasing upwards. The bottom spectra are for no
infall velocity, the middle spectra are for an infall velocity equal to the
velocity dispersion of 0.1~km~s$^{-1}$, and the top spectra are for an infall
velocity of 0.2~km~s$^{-1}$. The first thing we notice is that the
\hcop~(\jequals{1}{0}) lines are not very blue-asymmetric. Although the width
increases as the infall velocity in the core increases, the infall asymmetry
does not because the \jequals{1}{0}\ transition is mostly sensitive to motions
in the envelope. The emission from the core is mostly absorbed by the envelope
because the optical depth in the \jequals{1}{0}\ line is high. The
\hcop~(\jequals{3}{2}) transition, which is more optically
thin, does a better job of probing the motions in the core, and shows a greater
degree of asymmetry as the infall velocity increases.

\begin{figure}
\begin{center}
\resizebox{5in}{!}{\includegraphics{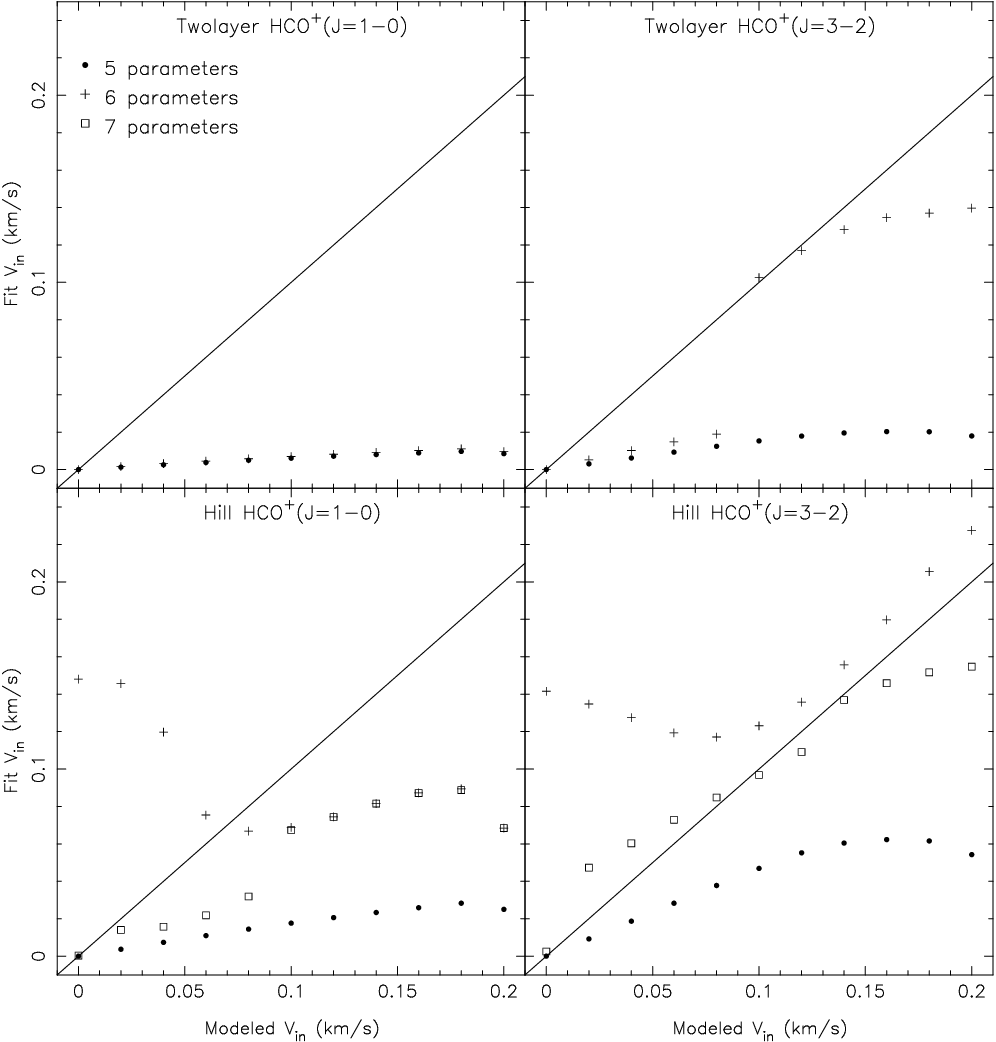}}
\end{center}
\caption{\label{mcf} Infall velocity fits. The above figures indicate the
infall velocities obtained by fitting {\sc twolayer5}, {\sc twolayer6}, {\sc
hill5}, {\sc hill6}, and {\sc hill7} models to the Monte
Carlo simulated
spectra in Simulation~C (Core Infall). The left hand panels are the results of
fitting the \hcop~(\jequals{1}{0}) spectra, and the right hand panels are the
results of fitting the \hcop~(\jequals{3}{2}) spectra. The top two figures
indicate the infall velocities obtained from the two-layer models. Below the
two-layer results are the results of the hill models. The symbols indicate the
number of free parameters, closed circles indicate the 5 free parameter models
({\sc twolayer5} and {\sc hill5}). Crosses indicate the 6 free parameter models
({\sc twolayer6} and {\sc hill6}). Open squares indicate the 7 free parameter
model ({\sc hill7}). The {\sc twolayer6} model often has two local minima, one
of which is usually fit by the {\sc twolayer5} model. We show the other local
minimum when it is available in this plot.}
\end{figure}

The low level of asymmetry in the \jequals{1}{0}\ line results in a relatively
poor fit to the infall velocity for both the two-layer model variants and the
hill model variants. Figure~\ref{mcf} depicts the efficacy of the model
variants at matching the simulated infall velocity in this simulation. The {\sc
twolayer5}, {\sc twolayer6}, {\sc hill5}, and {\sc hill7} all significantly
underestimate the infall velocity from the \hcop~(\jequals{1}{0}) line
profiles. This is not surprising as that transition is very thick and not very
sensitive to the motions in the core, where the infall is occurring. The {\sc
hill5} fit tends to be consistently a factor of $\sim5$ lower than the
simulated infall velocity. The situation is more encouraging when examining the
thinner \hcop~(\jequals{3}{2}) line profile fits. The {\sc twolayer5} model
continues to significantly underestimate the infall velocity, but the {\sc
twolayer6} model matches the simulated infall velocity well for velocities
between 0.1~km~s$^{-1}$ and
to 0.15~km~s$^{-1}$. Overall the RMS error on the infall velocity fits of the
{\sc twolayer6} model to the \hcop~(\jequals{3}{2}) lines is
0.03~km~s$^{-1}$. The {\sc hill5} model continues to underestimate the 
infall velocity from the \hcop~(\jequals{3}{2}) line, resulting in an RMS error
on the fit infall velocity of 0.07~km~s$^{-1}$. We believe this is due to
the fact that the red peak quickly becomes a shoulder in this simulation, and
the {\sc hill5} model tends to do much better when fitting a distinct red
peak. The {\sc hill6} model overestimates the infall velocity, while the {\sc
hill7} model tends to do about as well as the {\sc twolayer6} model at
estimating the infall velocity when
fitting the \hcop~(\jequals{3}{2}) lines in this simulation. Both the {\sc
hill7} model and the {\sc twolayer6} model begin to
underestimate the infall velocity for core infall velocities greater that
approximately 0.14~km~s$^{-1}$. At this point we run into the problem that
the line profile is not becoming
more asymmetric, just wider, due to the increasing velocities in the core.

\subsection{Simulation D --- Dense Bonnor-Ebert Sphere}
\label{besphere}
In our fourth simulation we choose a different density profile and a uniform
velocity throughout the cloud. We specifically choose the Bonnor-Ebert fit to
the density profile of B68 obtained by using extinction measurements of the
background K giant population to produce an azimuthally averaged column density
profile \citep*{all2}. We calculate the Bonnor-Ebert density profile using the
modified Lane-Emden equation
\begin{equation}
\frac{1}{\xi^{2}}\frac{d}{d\xi} \left(\xi^{2}\frac{d\Psi}{d\xi}\right) = e^{-\Psi},
\end{equation}
where $\xi = (r/a)\sqrt{4\pi G\rho_{c}}$ is the non-dimensional radial
parameter, $a$ is the isothermal sound speed, $\rho_{c}$ is the volume density
at the cloud center, and $\Psi(\xi) = -\ln(\rho/\rho_{c})$. We split this
second order differential equation in to two first order
equations by using the variable substitutions suggested by \citet*{bkv} and
integrate the equations numerically to derive the density profile. We use
$\xi_{\rm max} = 6.9$, the free parameter derived by \citet{all2} to be the
best fit to the B68 cloud extinction observations. \citet{dhwb} find that the
relative abundance of \hcop\ to molecular hydrogen is less than $1.4\times
10^{-10}$, which is significantly lower than the value of $2\times 10^{-9}$
which we have been using. Under such low abundances the \hcop\ lines we would
derive from radiative transfer calculations within this model cloud would be
optically thin. Since optically thick lines are required for our analysis, we
choose to maintain the \hcop\ to molecular hydrogen relative abundance at $2
\times 10^{-9}$ and only change the density profile in this simulation. We
divide the spherical cloud into thirty equally spaced shells, the central shell
has a molecular hydrogen density of $2.6\times 10^{5}$~cm$^{-3}$, while the
outermost shell has a molecular hydrogen density of $1.6\times
10^{4}$~cm$^{-3}$ and a maximum radius of $1.87\times 10^{17}$~cm or
12,500~AU. We assume a
constant kinetic temperature of 11~K throughout the cloud, as obtained by
\citet{lvlk} using NH$_{3}$ (1,1) and (2,2) hyperfine line observations
conducted at the DSN 34m telescope. 

\begin{figure}
\begin{center}
\resizebox{5in}{!}{\includegraphics{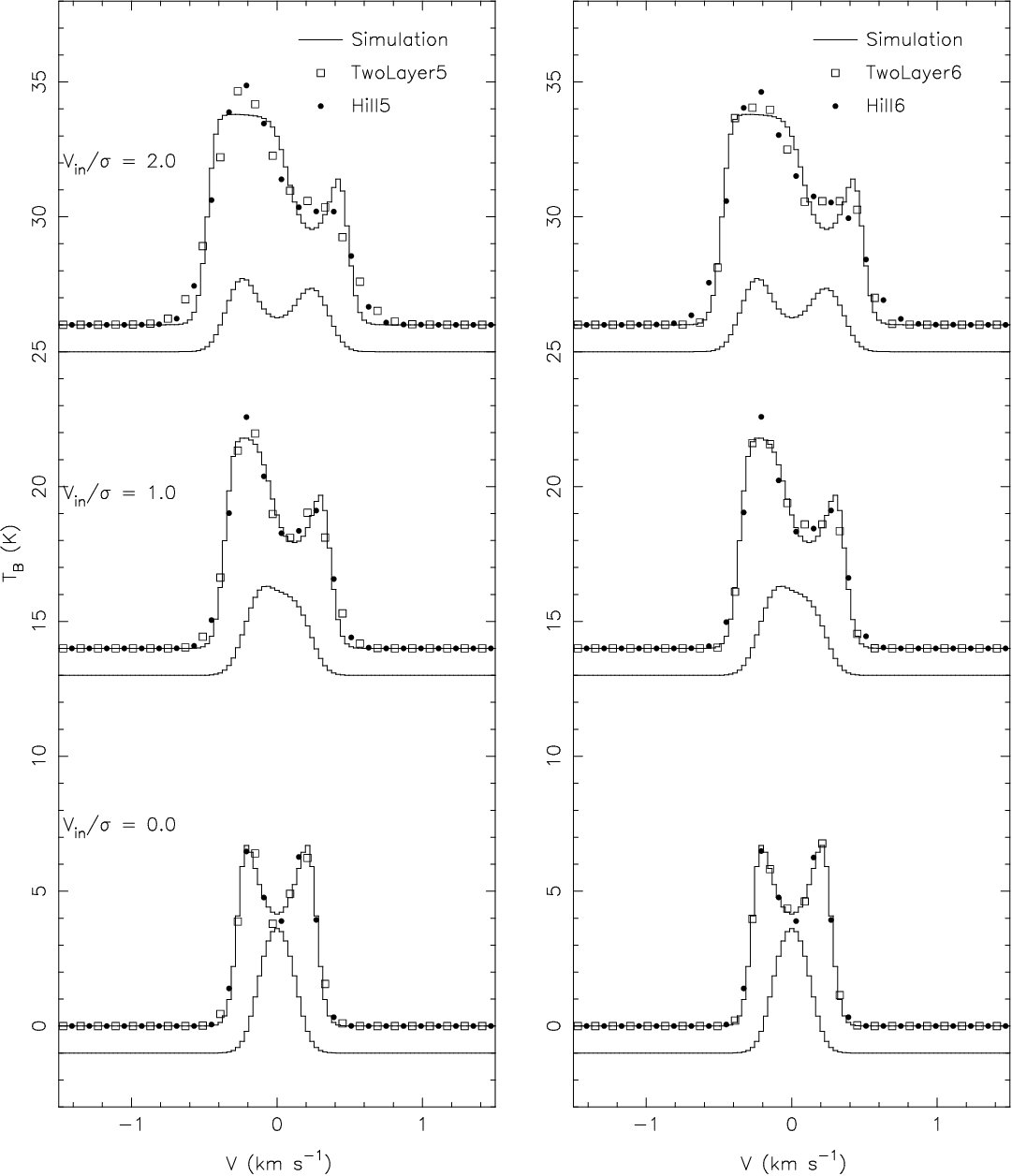}}
\end{center}
\caption{\label{mdjo} Monte Carlo simulated spectra (solid lines) of the
\hcop~(\jequals{1}{0}) and \hthcop~(\jequals{1}{0}) (multiplied by a factor of
2) emission from
Simulation~D (Dense Bonnor-Ebert Sphere). The infall velocity in the lowest set of spectra is
0~km~s$^{-1}$, those in the middle are of a cloud infalling at 0.12~km~s$^{-1}$,
while those at the top are of a cloud infalling at 0.24~km~s$^{-1}$. The
analytic model fits are overlaid on each \hcop\ spectrum. The open squares on
the left indicate the {\sc twolayer5} fits, while the filled circles are the
{\sc hill5} fits. One the right the open squares indicate the {\sc twolayer6}
fits, while the filled circles indicate the {\sc hill6} fit.}
\end{figure}

\begin{figure}
\begin{center}
\resizebox{5in}{!}{\includegraphics{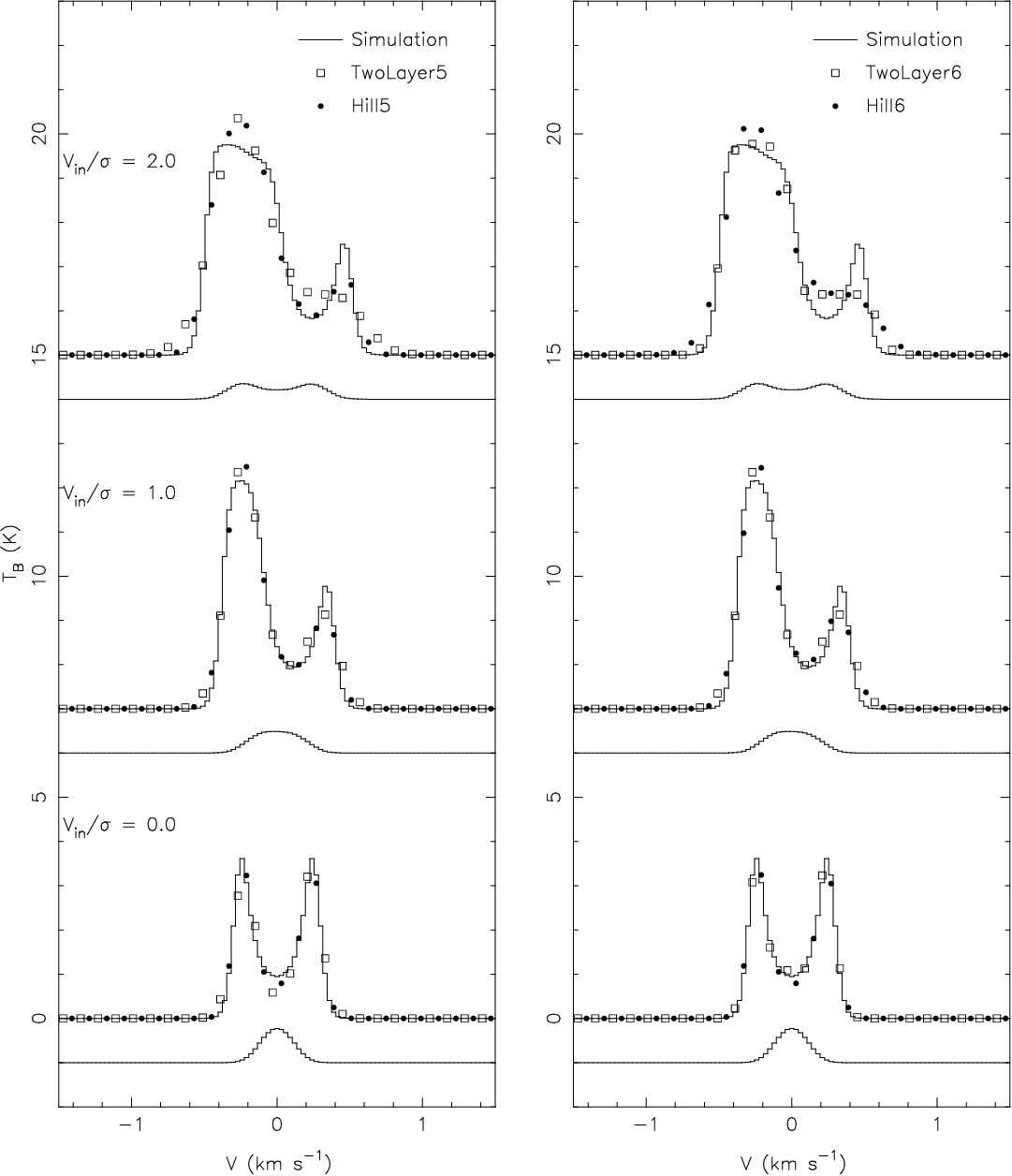}}
\end{center}
\caption{\label{mdjt} Monte Carlo simulated spectra (solid lines) of the
\hcop~(\jequals{3}{2}) and \hthcop~(\jequals{3}{2}) (multiplied by a factor of
2) emission from
Simulation~D (Dense Bonnor-Ebert Sphere). The infall velocity in the lowest set of spectra is
0~km~s$^{-1}$, those in the middle are of a cloud infalling at 0.12~km~s$^{-1}$,
while those at the top are of a cloud infalling at 0.24~km~s$^{-1}$. The
analytic model fits are overlaid on each \hcop\ spectrum. The open squares on
the left indicate the {\sc twolayer5} fits, while the filled circles are the
{\sc hill5} fits. One the right the open squares indicate the {\sc twolayer6}
fits, while the filled circles indicate the {\sc hill6} fit.}
\end{figure}

We assume a constant infall velocity throughout the cloud at all radii. We
modeled 11 infall velocities ranging from 0 to twice the velocity dispersion,
which we chose to be 0.12~km~s$^{-1}$. We base our choice of the velocity
dispersion on the \nthp~(\jequals{1}{0}) line width in B68 reported by
\citet{cbmt}. \hcop~(\jequals{1}{0}) and \hthcop~(\jequals{1}{0}) spectra from
this simulation are shown in figure~\ref{mdjo}. As in previous simulations we
choose a representative sample of three spectra in the direction of the cloud
center at different infall velocities. In this case the infall velocities are
0~km~s$^{-1}$, 0.12~km~s$^{-1}$, and 0.24~km~s$^{-1}$. On the left side of
figure~\ref{mdjo} the best fit {\sc twolayer5} and {\sc hill5} models are
shown, while on the right side the best fit {\sc twolayer6} and {\sc hill6}
models are shown. The first thing to note is that the \hthcop\ lines are
slightly asymmetric, similar to
\hthcop~(\jequals{1}{0}), HC$^{18}$O$^{+}$~(\jequals{1}{0}), and
D$^{13}$CO$^{+}$~(\jequals{2}{1}) observations of L1544 seen in figure~1 of
\citet{cwztdm1}. This is as a result of self-absorption, 
because although the \hthcop\ molecule has an abundance 64 times lower than
that of \hcop, it is 
not so low that it is optically thin. It is important to be aware of that
asymmetry may arise in less abundant species, however the degree of asymmetry
is indeed less than the asymmetry of the more abundant \hcop\ molecule. As in
previous simulations, the line profiles generated by the analytic models
reproduce the Monte Carlo generated profiles very well. Similar results for the
\hcop~(\jequals{3}{2}) and \hthcop~(\jequals{3}{2}) lines are shown in
figure~\ref{mdjt}. The \hthcop\ lines are more symmetric in this case as there
is less optical depth along the line of sight.

\begin{figure}
\begin{center}
\resizebox{5in}{!}{\includegraphics{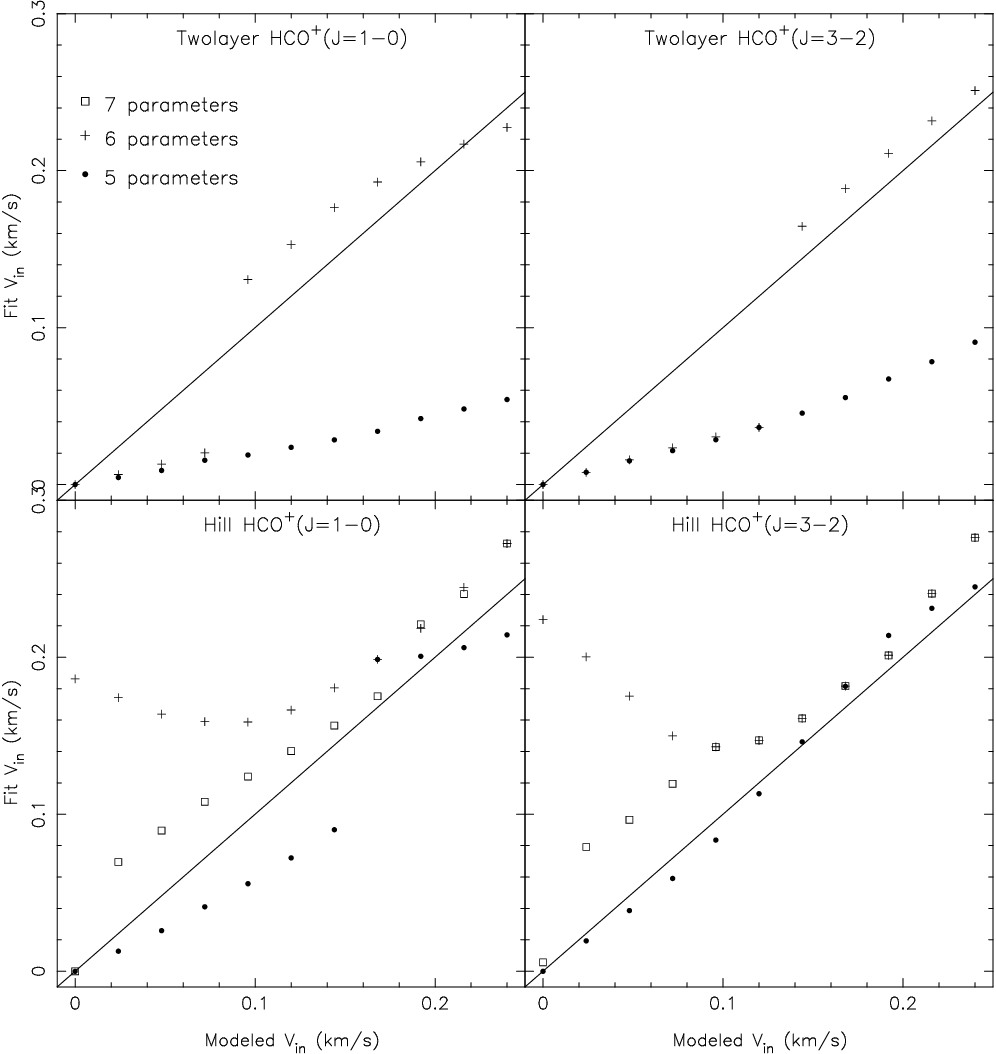}}
\end{center}
\caption{\label{mdf} Infall velocity fits. The above figures indicate the
infall velocities obtained by fitting {\sc twolayer5}, {\sc twolayer6}, {\sc
hill5}, {\sc hill6}, and {\sc hill7} models to the Monte
Carlo simulated
spectra in Simulation~D (Dense Bonnor-Ebert Sphere). The left hand panels are the results of
fitting the \hcop~(\jequals{1}{0}) spectra, and the right hand panels are the
results of fitting the \hcop~(\jequals{3}{2}) spectra. The top two figures
indicate the infall velocities obtained from the two-layer models. Below the
two-layer results are the results of the hill models. The symbols indicate the
number of free parameters, closed circles indicate the 5 free parameter models
({\sc twolayer5} and {\sc hill5}). Crosses indicate the 6 free parameter models
({\sc twolayer6} and {\sc hill6}). Open squares indicate the 7 free parameter
model ({\sc hill7}). The {\sc twolayer6} model often has two local minima, one
of which is usually fit by the {\sc twolayer5} model. We show the global
minimum in this plot.}
\end{figure}

We compare the infall velocities in this simulation with those obtained by
fitting the line profiles with our analytic models in figure~\ref{mdf}. As in
previous sections we present the fits to the \hcop~(\jequals{1}{0}) line
profiles on the left panels of the figure, and those to the
\hcop~(\jequals{3}{2}) line profiles on the right panels of the figure. The
upper panels refer to the two-layer model fits, while the lower panels refer to
the hill model fits. Neither of the two-layer models reproduce the simulated
infall velocity across the entire range of infall velocities we tested. The RMS
systematic error in the {\sc twolayer6} model fits of \hcop~(\jequals{1}{0}) is
0.03~km~s$^{-1}$, and 0.04~km~s$^{-1}$ with respect to the
\hcop~(\jequals{3}{2}) fits. The {\sc hill5} model does a better job at
estimating the infall velocity in the simulated clouds. The RMS systematic
error in the {\sc hill5} estimates of the infall velocity based on the
\hcop~(\jequals{1}{0}) profile fits is 0.03~km~s$^{-1}$, while the RMS
systematic error based on the \hcop~(\jequals{3}{2}) profile fits is
0.01~km~s$^{-1}$. As we have seen in previous simulations the {\sc twolayer6}
model yields a closed fit to the infall velocity at higher infall
velocities. The {\sc hill5} model provides a fairly good fit for all infall
velocities. These results are similar to the results obtained in Simulation~A
(Constant Infall),
which is the other simulation in which infall was constant at all radii. This
is reassuring as the central densities in these two simulations differ by a
factor of 4 and they have different radial density profiles. This suggests that
these analytic models may be able to predict infall velocities from optically
thick line profiles for a wide range of densities and radial density profiles
in star forming clouds.

\subsection{Error Analysis}
There are two components to the error in any fit of a model to data. The first,
systematic error, is error which arise in the model's description of the
data. It is due to the fact that the model may not correctly describe the
nature of the observed region. The second component is random error, which
arises from noise in the observation. Although random error can be reduced over
time or with better instrumentation, usually up to some limit, it is always a
factor in any observation, and in fitting a model to those observations. In the
preceding sections we examined how well our analytic radiative transfer models
fit synthetic noiseless spectra. From this we found systematic errors due to
the models shown in figures~\ref{maf} and~\ref{mbf}. In order to investigate
the behavior of the fits in the presence of random error we decided to add
noise to the spectra.

\citet{bg} have derived analytic expressions for the contribution of random
error in a spectrum to the three parameters of a Gaussian fit based upon the
work of \citet{lrdt}. They find that the random error in the line of sight
velocity and in the velocity dispersion of a Gaussian fit are the same and
equal to
\begin{equation}
\label{gauss}
\sigma_{v_{\rm LSR}} = \sigma_{\sigma} = 1.06\left(\delta_{v}\sigma
\right)^{1/2} \frac{\sigma_{\rm RMS}}{T_A^{*}},
\end{equation}
where $v_{\rm LSR}$ is the line of sight velocity of the Gaussian, $\sigma$ is
the velocity dispersion of the Gaussian, $\delta_{v}$ is the width of the
spectral channels, $T_A^{*}$ is the peak intensity of the Gaussian line
profile, and $\sigma_{\rm RMS}$ is the RMS noise level in the
spectrum. Unfortunately deriving expressions for the random error in the
parameters of the two-layer or hill models is more difficult. We therefore use
a bootstrap method to investigate the errors in the parameters of these models.
We generate a noiseless synthetic spectrum from one of the above simulations,
add random noise to it, and then perform the analytic radiative transfer model
fits. We do this 100 times with different random noise of the same RMS value in
order to record the variance of each model parameter. Assuming the errors are
normally distributed, we find that the random error in each parameter is merely
the square root of the variance in that parameter. 

\begin{figure}
\begin{center}
\resizebox{4in}{!}{\includegraphics{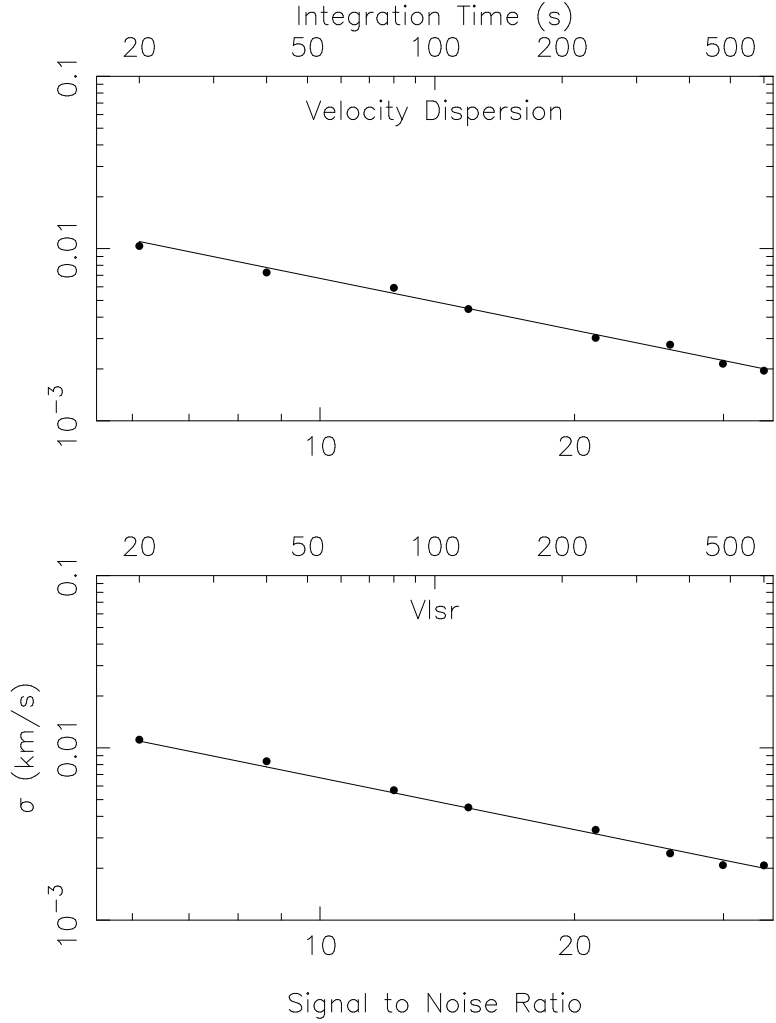}}
\end{center}
\caption{\label{gaussboot} Random errors in a Gaussian fit to the velocity dispersion and line of
sight velocity. We estimated errors to a Gaussian fit using a bootstrap method
and compared those errors to the theoretically predicted errors in a Gaussian
fit to a noise spectrum derived by \citet{bg}. The filled circles are our
bootstrap estimates for several different RMS noise levels. We use signal to
noise ratios in the lower x-axis. We also convert the RMS
to an integration time assuming a velocity resolution of 0.04~km~s$^{-1}$ in
the spectrum, a main beam efficiency of 0.5 and a system temperature of
200~K and place that integration time estimate on the upper x-axis. The
Gaussian we fit in the example has a peak line temperature of 5.01~K
and a velocity dispersion of 0.1~km~s$^{-1}$.} 
\end{figure}

We initially tested the method on Gaussian line fits, whose random errors
should follow the expression in equation~\ref{gauss}. Figure~\ref{gaussboot}
shows our resulting error determinations for different RMS noise values (filled
circles) versus the expected value (solid line). We express the RMS noise in
terms of the signal to noise ratio on the lower axis, and also in terms of the
integration time it would take to achieve a given RMS for a 
telescope with a system temperature of 200~K and a main beam efficiency of 0.5
when observing a Gaussian line with a peak brightness temperature of 5~K, along
the top axis. We
also use a channel width of 0.04~km~s$^{-1}$ in this case, less than the input
0.1~km~s$^{-1}$ velocity dispersion of the Gaussian line and typical of current
spectrometers. The bootstrap estimates are a good fit to equation~\ref{gauss},
with minimal scatter due to the random nature of the technique. This test
indicates that the bootstrap method is a robust estimator of the random error
in the parameters of the Gaussian fit and should be a robust estimator of the
random error in the parameters of the analytic radiative transfer model fits.

\begin{figure}
\begin{center}
\resizebox{3.8in}{!}{\includegraphics{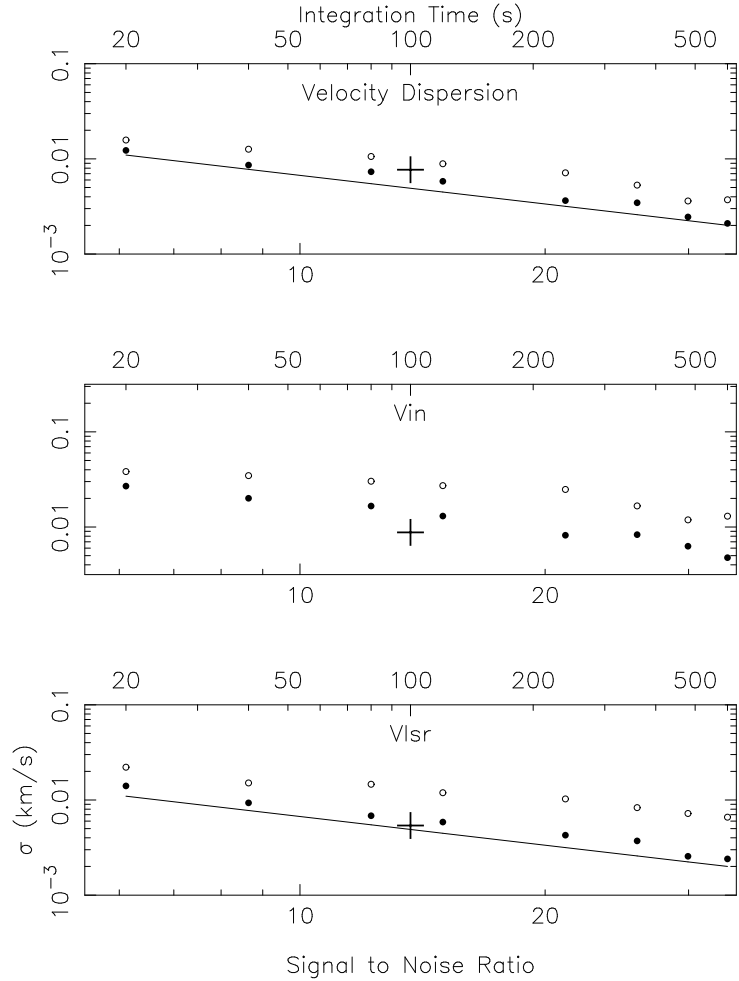}}
\end{center}
\caption{\label{hillboot} Systematic and random errors associated with the {\sc
hill5} model. We
measured the average systematic error in the dispersion velocity, infall
velocity, and line of sight velocity relative to our input models and plot them
above as the large crosses. We then added noise to simulate integration times
between 20 seconds and 600 seconds with velocity resolutions of 0.04
km~s$^{-1}$ (filled circles) and 0.16 km~s$^{-1}$ (open circles), and use the
bootstrap method to measure the random errors in those same parameters. The
upper x-axis indicates these integration times. The lower x-axis has a
measurement of the signal to noise ratio for the 0.04~km~s$^{-1}$ resolution
case. From
this it is possible to estimate the signal to noise ratio required in
order to have the random error drop below the systematic error for a given
observation. We find that signal to noise ratios greater than 30 are
required, using the \hcop~(\jequals{1}{0}) lines modeled in figure~\ref{majo}
as a template. The solid line indicates the theoretical Gaussian errors for
dispersion velocity and line of sight velocity as a function of time under the
same assumptions.}
\end{figure}

We apply the bootstrap method to the \hcop~(\jequals{1}{0}) line profiles of
Simulation~A (Constant Infall) shown
in figure~\ref{majo}. Figure~\ref{hillboot} summarizes our findings for the
random errors in the velocity dispersion, infall velocity, and line of sight
velocity parameters of the {\sc hill5} model. The closed circles indicate the
error in
parameters as a function of integration time for a spectral resolution of
0.04~km~s$^{-1}$, while the open circles are those for a spectral resolution of
0.16~km~s$^{-1}$. The error increases with the lower spectral resolution
because the self-absorption feature starts to be resolved out at that
resolution. The cross in the center of each panel indicates the average
systematic error obtained from the fits on figure~\ref{maf}. The average
systematic error in the velocity dispersion is 0.008~km~s$^{-1}$, the average
systematic error in the infall velocity is 0.009~km~s$^{-1}$, while the average
systematic error in the line of sight velocity is 0.005~km~s$^{-1}$. The solid
line in both the velocity dispersion panel and the line of sight velocity panel
are the Gaussian errors from figure~\ref{gaussboot}. Unsurprisingly the errors
in the hill model parameters are greater than those in the Gaussian fit, which
has fewer free parameters of the {\sc hill5} model, however they tend to
follow a similar power law trend with increasing integration times. What we can
see from this plot is that it is possible to be dominated by the systematic
errors of the hill model after a reasonable amount of integration time on a
typical millimeter radio telescope. For spectral resolutions adequate to
observe the line shape, the velocity dispersion random error drops below the
systematic level after less than 30 seconds of integration, while the line of
sight velocity is well determined after less than two minutes. Getting to a
point where the systematic errors in the infall velocity dominate over the
random errors takes the longest amount of time, but only about 10 minutes for a
line whose brightness temperature is approximately 5~K. This means that
generally the infall velocity is as well determined as it can get by the hill
model when the signal to noise ratio is greater than $\sim30$. We assume, based
on our experience with gaussian line fitting that maintaining a similar signal
to noise ratio in lines of differing peak intensities should yield similar
errors. This implies an integration time of 30 minutes for a line whose
brightness temperature is approximately 3~K, and 250 minutes for a line whose
brightness temperature is approximately 1~K. This means the observations
required to obtain the required signal to noise can take a prohibitively long
time for weak asymmetric line profiles.

\subsection{Depletion and Beam Smoothing}
\label{depletionbeam}
\begin{figure}
\begin{center}
\resizebox{5in}{!}{\includegraphics{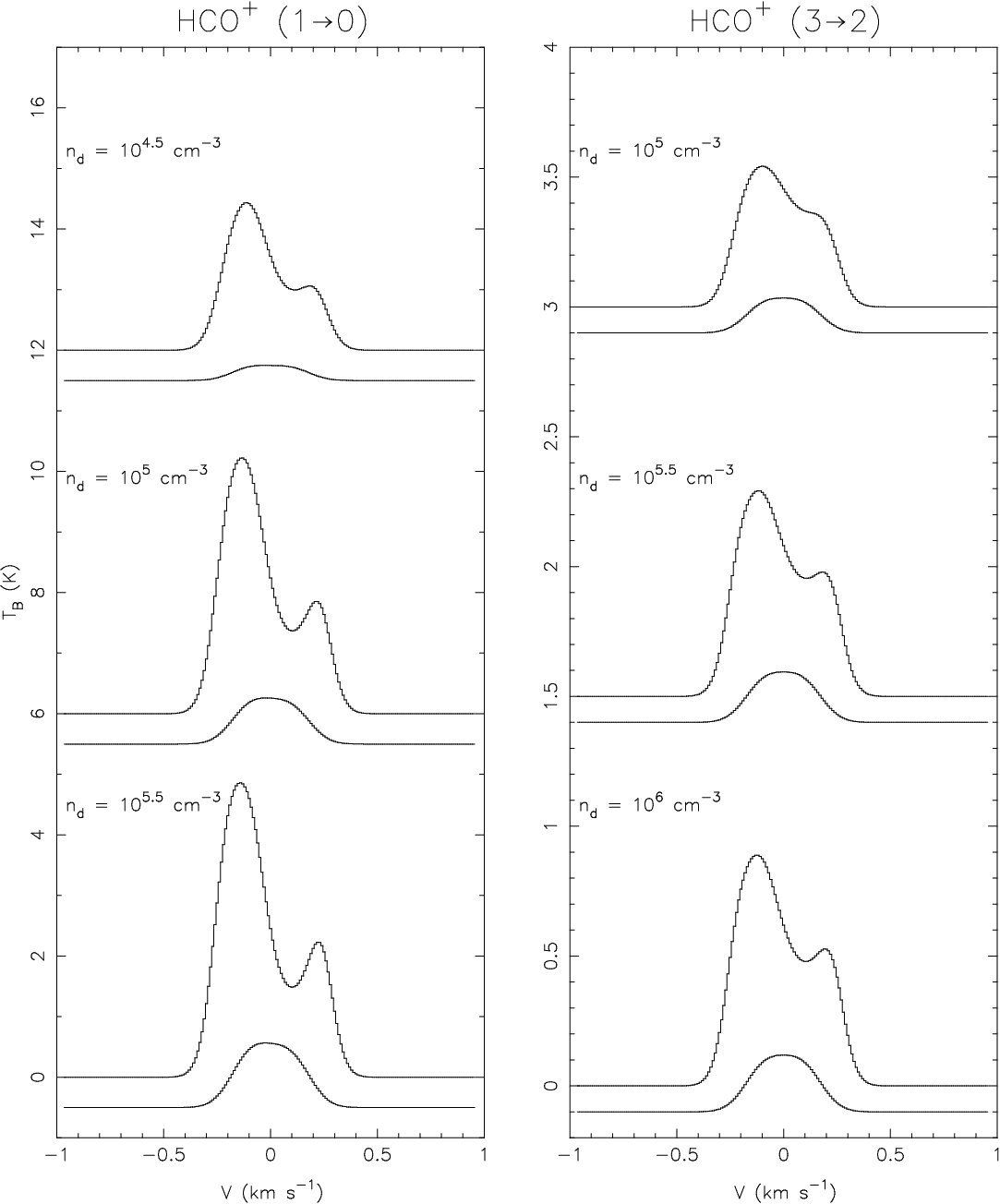}}
\end{center}
\caption{\label{depletion}The effect of depletion on the \hcop and
\hthcop~(\jequals{1}{0}) 
(left) as well as the \hcop and \hthcop~(\jequals{3}{2}) (right) line
shapes. The depletion increases as the depletion density ($n_{d}$) drops. In
each column we highlight the range of depletions in which the transition
remains bright ($T_{r} > \sim 1$ K). In most cases the {\sc hill5} model
continues to provide a good fit to the infall velocity. The only exception is
the upper-most set of spectra in which the {\sc hill5} model is not applicable
according to the prescription given in \S\ref{method} as the dip between the
peaks is not sufficiently deep. The \hthcop~(\jequals{1}{0}) spectra have been
multiplied by a factor of 5, and the 
\hthcop~(\jequals{3}{2}) spectra have been multiplied by a factor of 20.}
\end{figure}

Depletion in high density regions of a molecular cloud core is a significant
effect in observed infall regions. Many observations of CO, CS, and \hcop\ in
dense cores show a ring
distribution thought to result from an abundance drop in the gas phase due to
freeze out of these molecules onto grains
\citep{klv,wlv,kallsuw,all,cwtdm,bclal,jw,blccsl,tmcwc,tmcw}. We have
added depletion in some cases to perform an investigation of the
efficacy of the analytic models at estimating the infall velocity as depletion
increases. We adopt the formalism of \citet{tmcwc} and set the abundance
of \hcop\ to vary exponentially with the density,
\begin{equation}
X(r) = X_{0} \exp\left[-n(r)/n_{d}\right],
\end{equation}
where $X(r)$ is the abundance of \hcop\ as a function of cloud radius, $X_{0}$
is the \hcop\ abundance in the low density limit ($2\times 10^{-9}$), $n(r)$ is
the density of molecular hydrogen as a function of cloud radius, and $n_{d}$ is
the density at which the abundance drops by a factor of $e$. We included
depletion in the
Simulation~A (Constant Infall) case where $v_{\rm in} = 0.1$ km~s$^{-1}$ for
$n_{d} = 10^{6},
10^{5.5}, 10^{5}, 10^{4.5}, 10^{4}, 10^{3.5},$ and $10^{3}$ cm$^{-3}$. We found
for the {\sc hill5} model the fit infall velocity changed by less that one
hundredth of a kilometer per second down to an $n_{d}$ of $10^{4}$ cm$^{-3}$
for the \hcop~(\jequals{1}{0}) line and $10^{5}$ for the
\hcop~(\jequals{3}{2}) line. Figure~\ref{depletion} illustrates that in each of
these cases that level of depletion
corresponds to a change in the line profile from a two-peaked self-absorbed
profile to a ``shoulder'' profile. Severe depletion, beyond the levels reported
in figure~\ref{depletion} tends to result
in a line too weak to be observed. Based on this simulation, we infer that
moderate levels of depletion which does not render the line profile symmetric
or unobservable has
only a small effect on the {\sc hill5} model's
infall velocity determinations.

\begin{figure}
\begin{center}
\resizebox{5in}{!}{\includegraphics{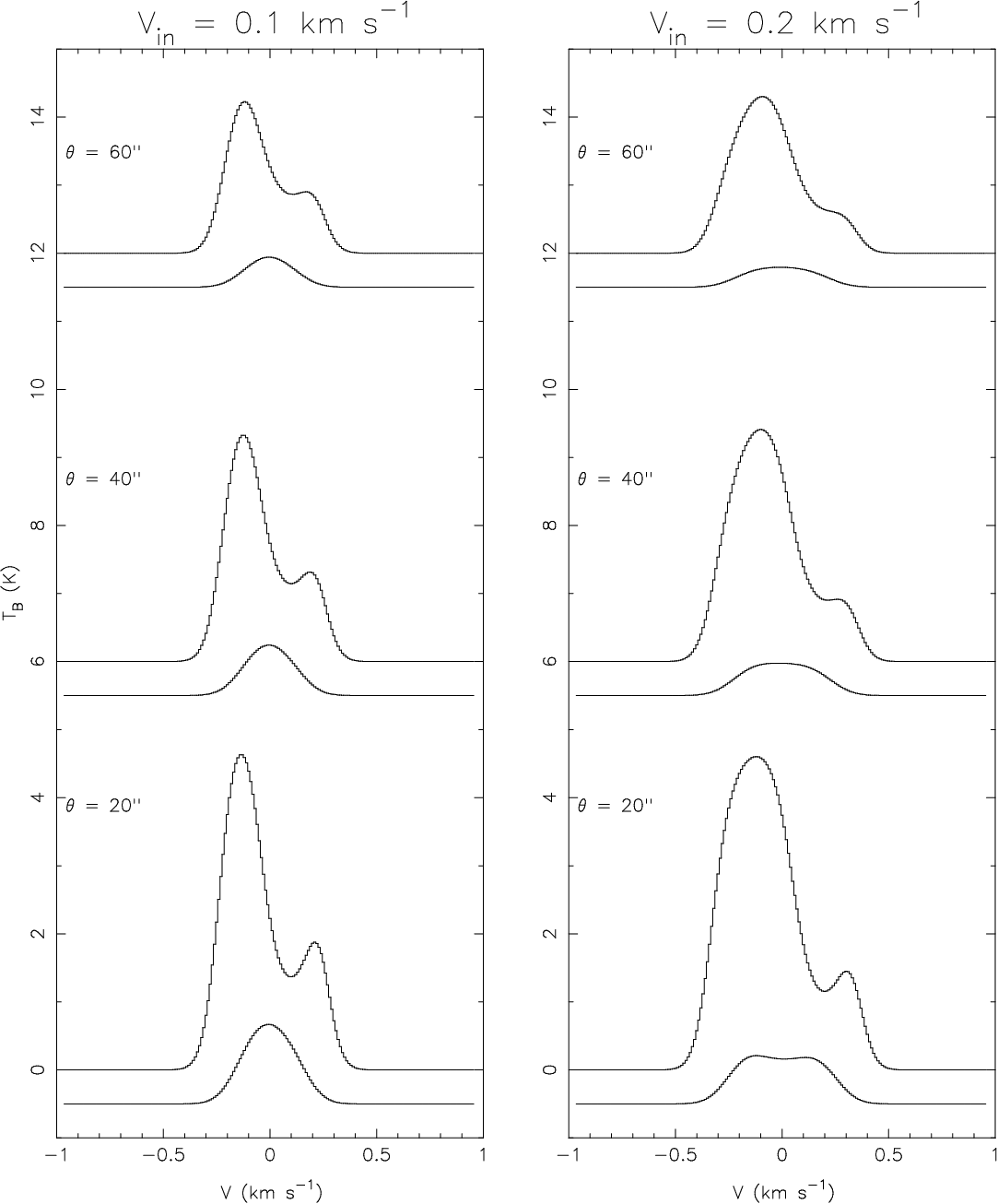}}
\end{center}
\caption{\label{beam}The effect of beam size of the \hcop~(\jequals{1}{0}) and
\hthcop~(\jequals{1}{0}) line profiles for Simulation~A (Constant Infall) clouds with infall
velocities of 0.1 and 0.2 km~s$^{-1}$. The simulated cloud is placed at a
distance of 140~pc from the observer, typical of the nearest starless cores,
and convolved with various beam sizes typical of current millimeter
telescopes. The lowest spectra represent a beam size of 20\sec, the middle
spectra represent a beam size of 40\sec, while the uppermost spectra represent
a beam size of 60\sec. The left panel is a simulation of  an infall velocity of
0.1~km~s$^{-1}$ and the right panel is a simulation of an infall velocity of
0.2~km~s$^{-1}$. The \hthcop~(\jequals{1}{0}) spectra have been multiplied by a
factor of 5.}
\end{figure}

Another important observational effect is smoothing by a telescope beam. Our
simulated line profiles were generated by
integrating along a pencil beam through the center of a simulated cloud. We
have generated 2 dimensional grids of pencil beam integrations of our simulated
clouds and can convolve them to produce a simulated Gaussian telescope beam. We
begin by projecting the cloud on the sky plane at a distance of 140~pc, so that
a beam of 60\sec\ is equal to a distance of about 0.04~pc in the cloud. We then
convolved two cases from Simulation~A (Constant Infall), where $v_{\rm in} =
0.1$ and $0.2$ 
km~s$^{-1}$, to generate beam sizes of 20\sec\ (0.014~pc), 40\sec\ (0.027~pc),
and 60\sec\ (0.041~pc) in the
\hcop~(\jequals{1}{0}) and \hthcop~(\jequals{1}{0}) lines. The resulting
beam-smoothed line profile seen toward the center of each simulated clouds
is shown in figure~\ref{beam}. From the left column of figure~\ref{beam} we see
that beam smoothing has a similar effect to depleting the core. This is not
surprising as both depletion and beam smoothing tend to increase the relative
contribution of emission from the envelope of the molecular cloud. Our
investigation indicates that the infall speed is well fit in cases
where the beam width is less than or equal to the core diameter ($10^{17}$~cm
in our simulation), however this is dependent on both the density structure and
velocity structure of the molecular cloud. In practical terms, a higher
resolution probes a smaller region which may have a large infall velocity and
therefore a more asymmetric line.

\section{Applying the Analytic Models}
\label{method}
As a final exercise we have applied the {\sc hill5} model to a high
signal to noise line profile. We have learned from the
simulations that the {\sc hill5} model is the most robust model for
determining the infall velocity of a molecular cloud from the line
profile for a majority of line profile shapes. In particular the {\sc hill5}
model yields a good fit when the line profile has distinct blue-shifted and
red-shifted peaks with a self-absorption minimum between them. For line
profiles with this shape, in simulations A and D, which had the same infall
speed at all radii, the RMS error of the {\sc hill5} infall velocity
determination is 0.01~km~s$^{-1}$, better than the overall RMS error in these
simulations of 0.02~km~s$^{-1}$. We recommend using the {\sc hill5} model to
derive infall velocities from asymmetric molecular line profiles.

In some cases there is little or no red-shifted peak, but merely a shoulder on
the red side of the line profile. These cases tend to occur when the infall
velocity is greater than the velocity dispersion. In Simulation~A (Constant
Infall) and Simulation~C (Core Infall) red-shoulder profiles occur in the
\hcop~(\jequals{3}{2}) transition for simulated infall velocities greater than
0.1~km~s$^{-1}$, in Simulation~B
(Envelope Infall) red-shoulder profiles occur in both the transitions for
infall velocities greater than 0.1~km~s$^{-1}$, while in Simulation~D (Dense
Bonnor-Ebert Sphere) red-shoulder profiles never occur for all the infall
velocities in our simulations. In these ``shoulder'' spectra the {\sc
twolayer6} model
gives a comparable infall velocity estimate to the {\sc hill5}
model. We note that in Simulation~C, where the {\sc twolayer6} model performs
better than the {\sc hill5} model the red-shoulder model, lines tend to be weak
($<1$~K). When using the {\sc twolayer6} model it is important to choose the
local minimum (which should also be
the global minimum in these cases) that corresponds to the ``shoulder''
solution set. One can verify that the ``shoulder'' solution has been found by
examining the best fit excitation temperature of the front layer, which should
be greater than the background temperature. If the front layer excitation
temperature is equal to the background temperature then it is likely that the
one has found the ``dip'' solution which is usually never a good fit to the
infall velocity.

Why is the {\sc hill5} model good at finding the infall velocity of most
clouds, while the {\sc twolayer6} model is successful only on spectra with an
asymmetric shoulder? The key to understanding this lies in how well each model
matches the excitation conditions along the line of sight at each Doppler
velocity. In figure~\ref{excivtau} we presented cases in which both the {\sc
hill5} and {\sc twolayer6} models match the excitation profile of the
simulation quite well for $\tau<3$ at the line center velocity. Although the
figure depicts only one Doppler velocity, the match continues to hold over the
entire line profile. The resulting infall velocities derived from these models
closely match the infall velocities in the simulated contracting cloud. In
cases where the infall velocities derived from the analytic models do not match
the infall velocities in the simulated clouds, the modeled excitation
temperature profiles as a function of optical depth along the line of sight
derived from the spectral line fits do not match the simulated excitation
temperature profiles. Even in cases where the spectral line match is quite
good, if the excitation conditions are not well matched then all the physical
parameters, including the infall velocity, will not match the parameters of the
simulated or observed cloud. 

We find that the {\sc hill5} model tends to reproduce the excitation
temperature profile as a function of optical depth quite well for a wide range
of infall velocities. The reason that this is true is that the excitation
conditions in a cloud do tend to rise as a function of optical depth over
the first couple of optical depths, until the excitation temperature reaches
the kinetic temperature of the core or begins dropping again as the density
drops. In all of our simulations a linear fit to the rising excitation
temperature, starting with an excitation temperature equal to the background
temperature, is a good approximation of the actual excitation temperature
profile at all Doppler velocities. As a result the {\sc hill5} model is usually
sensitive to the physical conditions of the collapsing cloud and yields a good
estimate of the infall velocity if the cloud radii which comprise the first few
optical depths of the line profile are infalling. For this reason we recommend
using the {\sc hill5} model. Our simulations and error analysis suggest that
the line profile should have a peak brightness temperature greater than 1~K and
a signal to noise ratio greater than 30. The line profile must also have a well
defined dip or shoulder which must be resolved by the spectrometer
resolution. If these conditions are met the {\sc hill5} model can be expected
to provide an estimate of the infall velocity with an accuracy of
0.02~km~s$^{-1}$. Our simulations suggest the accuracy for dip profiles alone
is 0.01~km~s$^{-1}$. 

The {\sc twolayer6} model tends to reproduce the excitation temperature profile
as a function of optical depth well only for higher infall velocities which
tend to produce spectral lines that are very asymmetric or have well defined
red shoulders. In these cases it may be tempting to use the {\sc twolayer6}
method to derive the infall velocity, however there are several caveats which
one must take in to account. In cases with low infall velocities or more
symmetric line
profiles the excitation temperature of the front layer is often low (at or near
the cosmic microwave background temperature) while that of the rear layers is
often much higher than the maximum excitation temperature in the cloud, even in
cases where the {\sc twolayer6} model produces a good fit to the emerging line
profiles. In these cases the infall velocities as other physical conditions
derived from the {\sc twolayer6} model do not match the conditions in the
simulated cloud. The reason this occurs is that there is a certain amount of
degeneracy in reproducing a symmetric or nearly symmetric line profile with the
{\sc twolayer6} model. A very high excitation temperature region behind a
very low excitation temperature region produces the same nearly symmetric
profile as a moderately high excitation temperature region behind a moderately
low excitation temperature region. At higher infall velocities, when the line
profile is very asymmetric or has only a shoulder, the excitation temperature
of the front region can be accurately determined from the red-shifted emission
of that line profile. This breaks the degeneracy and creates an excitation
temperature profile along the line of sight that more closely matches the
simulation and is more well determined overall. As a result the physical
parameters of the simulation, including the infall velocity, tend to be well
determined by the {\sc twolayer6} model in these cases. 

\begin{figure}
\begin{center}
\resizebox{5.5in}{!}{\includegraphics{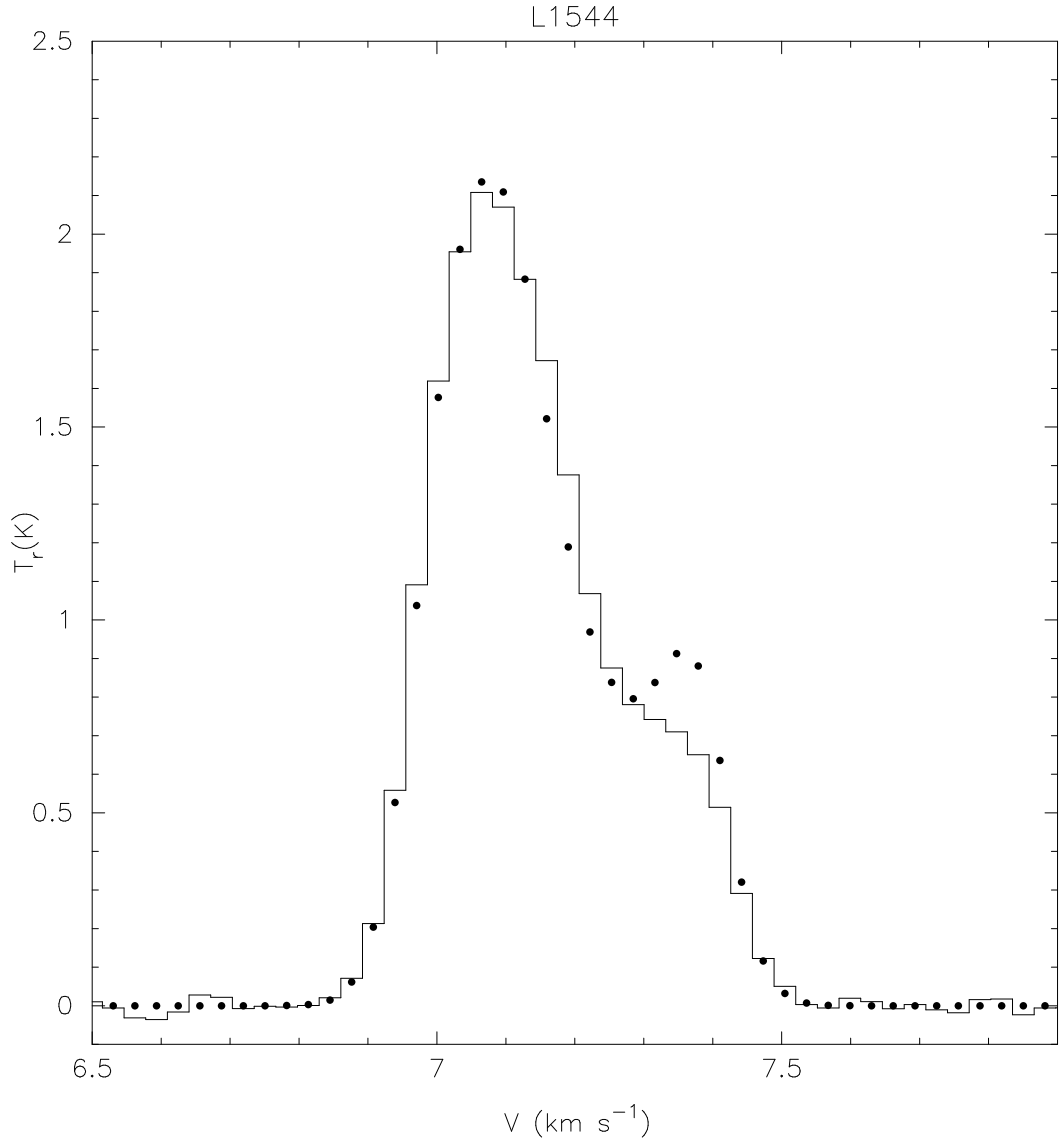}}
\end{center}
\caption{\label{data} Fits of the {\sc hill5} model to a
high signal to noise \nthp\ observation of L1544. The 
spectrum shown is the isolated \nthp~(\jequals{1}{0}) component observed by
Bourke et 
al. (in preparation) toward L1544. The filled circles indicates the best {\sc
  hill5} fit.}
\end{figure}

In figure~\ref{data} we present a high signal to noise line profiles upon
which we will carry out the procedure prescribed above. The line
profile is the isolated component of an 
\nthp~(\jequals{1}{0}) line profile in L1544 observed by Bourke et al. (in
preparation). Based on our results in the previous sections we choose to model
the emission using the {\sc hill5} model. That
best fit is shown by the points in figure~\ref{data}. We see that the fit is
fairly good, though it deviates from the line profile by producing a second
peak instead of a shoulder. The {\sc hill5} parameters that result in this fit
are shown in table~\ref{datahillfit}. In this case our {\sc hill5} infall
estimate of 0.099~km~s$^{-1}$ matches well with the best fit {\sc twolayer6}
estimate of 0.095~km~s$^{-1}$. Unfortunately since we did not
create this profile from 
a model we do not know the actual kinematic conditions in the cloud, however
there are two other estimates obtained using detailed radiative transfer
methods. \citet{krbp} have fit a model to \nthp~(\jequals{1}{0}) and
\nthp~(\jequals{3}{2}) spectra at 4 positions around L1544. They find an infall
speed profile that accelerates inwards from the cloud edge to
0.24~km~s$^{-1}$. Although this peak infall speed is much more than our
estimate, we note that their fits to L1544 include no velocity dispersion. We
speculate that the infall speed may be artificially high in order to match the
line width in the absence of any turbulent velocity component. Bourke et
al. (in preparation) have observed this particular source in several
molecular line transitions, include \nthp~(\jequals{1}{0}),
\nthp~(\jequals{3}{2}), DCO$^{+}$~(\jequals{2}{1}), DCO$^{+}$~(\jequals{3}{2}),
CS~(\jequals{2}{1}), CS~(\jequals{3}{2}), CS~(\jequals{5}{4}), and
C$^{34}$S~(\jequals{2}{1}). They have used the density model of this cloud
proposed by \citet{tmcwc} and assumed infall at the same velocity over the
entire cloud along with a kinetic temperature of 10 K to create radiative
transfer models of L1544 and match their observed line profiles. A simultaneous
match of all the components in the \nthp\ spectral lines currently yields an
infall velocity of 0.1 km~s$^{-1}$, which is in agreement with our estimate
to within a hundredth of a kilometer per second.

\begin{deluxetable}{lllll}
\tablecaption{\label{datahillfit}Parameters of best {\sc hill5} fit to L1544
  \nthp\ profile}
\tablecolumns{5}
\tablewidth{0pt}
\tablehead{\colhead{$\tau_{0}$} & \colhead{$v_{\rm LSR}$} & \colhead{$v_{\rm
  in}$} & \colhead{$\sigma$} & \colhead{$T_{P}$} \\ & \colhead{(km~s$^{-1}$)} &
  \colhead{(km~s$^{-1}$)} & \colhead{(km~s$^{-1}$)} & \colhead{(K)}}
\startdata
3.26 & 7.183 & 0.099 & 0.070 & 6.14 \\
\enddata
\end{deluxetable}

\section{Conclusion}
We have shown that analytic radiative transfer models are effective at
estimating the infall velocities of contracting clouds from the
blue-asymmetric line profiles of species of moderate optical depth in the
region of collapse. The models provide good velocity estimates with few
assumptions. The greatest advantage
of the analytic models is that they allow for a quick and reasonable estimate
of infall velocity with a minimum of computational effort. In cases where
nothing else is known about the source, they provide the only possible estimate
of infall, and in regions where a more robust model can be constructed, the
analytic models provide an excellent starting point for further
modeling \footnote{Implementations of the models used for this paper are
available from the authors at
\url{http://cfa-www.harvard.edu/\~{}cdevries/analytic\_infall.html}}.
The main conclusions of this work are as follows:

\begin{enumerate}
\item Analytic radiative transfer models can reproduce
the blue-asymmetric line profiles seen in infalling starless cores. In addition
to the two-layer model first presented in \citet{mmtww} we have found a second
class of radiative transfer models (the hill models) and explored 5 variants of
these two classes of analytic radiative transfer models. All the model
variations were able to produce blue-asymmetric line profiles similar to those
we observe in starless cores.

\item Successful analytic radiative transfer models reproduce not only the line
profile shape, but also match the initial excitation conditions along the line
of sight up to or beyond an optical depth of 1. When this requirement is met,
the modeled infall velocity
successfully fits the simulated infall velocity. Those models which match the
line profile
shape, but do not match the line of sight excitation conditions
cannot reveal the infall velocity and velocity dispersion in the cloud.

\item The {\sc hill5} analytic radiative transfer model is the preferred model
when fitting a blue-asymmetric line profile. In simulations with
constant velocity as at all cloud radii the
infall
velocities attained by fitting the {\sc hill5} analytic radiative transfer
model to the simulated line profiles match the infall speed with an RMS error
0.02~km~s$^{-1}$. The {\sc hill5} model tends to perform especially
well when there are two distinct peaks such that
the intensity difference between the self-absorption trough and the redshifted
peak is
greater than 10\% of the intensity of the blueshifted peak. The RMS error in
the infall determination of {\sc hill5} fits to simulated spectral lines with
those characteristics from Simulations~A and D is 0.01~km~s$^{-1}$.

\item The physical parameters of simulated clouds that produce blue-asymmetric
line profiles with no distinct shoulder or redshifted peak are generally not
well
determined by any of the model variations we have studied in this paper. We
recommend not using these models to estimate infall velocities from line
profiles that do not have a separate redshifted peak or a distinct redshifted
shoulder.

\item A better match to the line profile shape does not necessarily imply a
better match to the excitation profile or the physical parameters, such as the
infall velocity, of the observed cloud. This can be seen most clearly in
figure~\ref{dipshoulder} where local minimum of the ``dip'' solution often
provides 
a lower $\chi^{2}$ fit to the line profile shape, but a poorer fit to the
infall velocity than the ``shoulder'' solution. Following the instructions in
\S\ref{method} when fitting a blue-asymmetric line profile will minimize the
risk of obtaining an unusually poor estimate of the physical parameters from
the analytic radiative transfer model fit to the line profile.

\item All the analytic models lose sensitivity to the infall velocity
if there is a significantly optically thick static envelope between the
observer and the infall region. In these cases a more optically thin line can
provide a better estimate to the infall velocity in the obscured
region. Conversely an optically thin tracer may underestimate the infall
velocity of an infalling envelope around a static cloud core. These methods are
sensitive to the velocities and conditions in the regions in which the line is
becoming optically thick.

\item A peak signal to noise ratio greater than or equal to 30 is
required for the systematic error on the infall velocity estimate to be greater
than the random error. We recommend achieving a peak signal to noise ratio of
30 in order to minimize the random error of a line profile fit using our
analytic radiative transfer models, however nothing is gained by achieving a
higher signal to noise level. Achieving this signal to noise level can take a
prohibitively long time for lines whose peak antenna temperature is less than
or equal to 1~K.

\item Depletion and beam smoothing can greatly affect the observed
line profile in infalling starless cores. We conclude that, even in the
presence of these effects, if the line profile
is asymmetric and a good candidate to be fit using these analytic radiative
transfer models (as described in \S\ref{method}), then generally the radiative
transfer models do yield an accurate estimate of the infall
velocity. If the cloud is depleted to such an extent or viewed with such a
large beam that the line profile becomes very weak or symmetric, then these
models will not accurately probe the infall region of the starless core.
\end{enumerate}

\acknowledgements
We are grateful to Tyler Bourke for contributing his L1544 data and some
preliminary results from his radiative transfer modeling for use in this
paper. We also acknowledge Henrik Beuther for his encouraging advice and an
anonymous referee for his or her valuable comments and suggestions. This
research was conducted with support from NASA Origins of Solar Systems Grant
NAG5-13050.

\end{document}